%% file: main.tex
\documentclass{acmconf}
\usepackage{graphicx}
\usepackage{latexsym}
\usepackage{amssymb}

\title{Typestate Checking and Regular Graph Constraints}

\author{Viktor Kuncak and Martin Rinard\\
        Laboratory for Computer Science\\
        Massachusetts Institute of Technology\\ 
        Cambridge, MA 02139 \\ 
        {\tt $\{$vkuncak,rinard$\}$@lcs.mit.edu} \\[4ex]
        {\sf MIT-LCS-TR-863, September 2002}}

\begin{document}

\maketitle

\renewcommand{\thefootnote}{\fnsymbol{footnote}}
\footnotetext[1]{
This research was supported in part by DARPA Contract
F33615-00-C-1692, NSF Grant CCR00-86154,
NSF Grant CCR00-63513, and the Singapore-MIT Alliance.
}

\renewcommand{\thefootnote}{\arabic{footnote}}

\input{defs}

\input{abstract}


\input{introduction}
\input{definitions}
\input{implication}
\input{related}
\input{conclusion}
\input{acknowledge}

\bibliographystyle{plain}
\bibliography{pnew}

\end{document}


%% file: defs.tex
\newcommand{\qqquad}{\qquad\qquad\qquad}
\def\allowspace{\hskip 0em plus 1em minus 0em\relax}
\newcommand{\glngt}{1.5ex}
\newcommand{\glngts}{1ex}
\newcommand{\mnl}{\\[\glngt]}
\newcommand{\mnls}{\\[\glngts]}
\newtheorem{definition}{Definition}
\newtheorem{exa}[definition]{Example}
\newenvironment{example}{\begin{exa}\rm}
                        {\end{exa}$\eoex$\ \\[1ex]}
\newtheorem{property}[definition]{Property}
\newtheorem{lemma}[definition]{Lemma}
\newtheorem{proposition}[definition]{Proposition}
\newtheorem{theorem}[definition]{Theorem}
\newtheorem{claim}[definition]{Claim}

\newenvironment{proof}
  {\noindent {\bf Proof.}\allowspace}
  {\ifvmode\eop\vskip0em plus3em\goodbreak
  \else\eop\fi\medskip}

\newcommand{\eop}{\vrule width4pt height4pt depth0pt}
\newcommand{\eoex}{\blacklozenge}

\newcommand{\tu}[1]{\langle #1 \rangle}

\newcommand{\lefti}{[ \! [}
\newcommand{\righti}{] \! ]}
\newcommand{\tr}[1]{\lefti #1 \righti}
\newcommand{\str}[1]{\mathop{st}\lefti #1 \righti}
\newcommand{\ctr}[1]{\mathop{con}\lefti #1 \righti}

\newcommand{\m}[1]{\mbox{\sf #1}}

\newcommand{\graphimplies}{\leadsto} 
\newcommand{\graphequiv}{\approx} 


%% file: abstract.tex
\begin{abstract}
We introduce {\em regular graph constraints} and explore
their decidability properties.  The motivation for regular
graph constraints is 1) type checking of changing types of
objects in the presence of linked data structures, 2) shape
analysis techniques, and 3) generalization of similar
constraints over trees and grids.

Typestate checking for recursive and potentially cyclic data
structures requires verifying the validity of implication
for regular graph constraints.  The implication of regular
graph constraints also arises in shape analysis algorithms
such as role-analysis and some analyses based on
three-valued logic.  

Over the class of lists regular graph constraints reduce to
a nondeterministic finite state automaton as a special case.
Over the class of trees the constraints reduce to a
nondeterministic top-down tree automaton, and over the class
of grids our constraints reduce to domino system and tiling
problems.

We define a subclass of graphs called {\em heaps} as an
abstraction of the data structures that a program constructs
during its execution.  We show that satisfiability of
regular graph constraints over the class of heaps is
decidable.  However, determining the validity of implication
for regular graph constraints over the class of heaps is
undecidable.  The undecidability of implication is the
central result of the paper.  The result is somewhat
surprising because our simple constraints are strictly less
expressive than existential monadic second-order logic over
graphs.  In the key step of our proof we introduce the class
of {\em corresponder graphs} which mimic solutions of Post
correspondence problem instances.  We show undecidability by
exhibiting a characterization of corresponder graphs in
terms of presence and absence of homomorphisms to a finite
number of fixed graphs.

The undecidability of implication of regular graph
constraints implies that there is no algorithm that will
verify that procedure preconditions are met or that the
invariants are maintained when these properties are
expressed in any specification language at least as
expressive as regular graph constraints.
\end{abstract}


%% file: introduction.tex
\section{Introduction}

Types capture important properties of objects in the
program.  In an imperative language properties of objects
change over time.  It is therefore desirable that types
capture changing properties of objects.  A typestate system
is a system where types of objects change over time.  A
simple typestate system was introduced in
\cite{StromYemini86Typestate}, more recent examples include
\cite{KuncakETAL02RoleAnalysis,SmithETAL00AliasTypes,
WalkerMorrisett00AliasTypesRecursive,
DeLineFahndrich01EnforcingHighLevelProtocols}.  We view
typestate as a step towards statically checking properties
of objects \cite{LeinoStata97CheckingObjectInvariants,
FlanaganETAL02ExtendedStaticCheckingJava}.

One of the difficulties with defining object properties in
object-oriented languages is that a property of an object
may depend on properties of other objects in the heap.  Some
systems allow programmers to identify properties of an
object $x$ in terms of properties of objects $y$ such that
$x$ references $y$.  The idea that important properties of
an object $x$ may depend on properties of objects $z$ such
that $z$ references $x$ was introduced in the role
system~\cite{KuncakETAL02RoleAnalysis}.

In general, properties of objects may be mutually recursive
and the referencing graph of objects may be cyclic.  Due to
cycles in the heap the least fixpoint solution for the
recursive object properties is not acceptable because there
is no basis to ground the inductive definitions of these
properties.  We therefore say that a heap satisfies a set of
properties if there {\em exists} some choice of predicates
that satisfy the mutually recursive definitions.  The
existential quantification over predicates leads to
constraints that have the form of existential monadic
second-order logic
\cite{FaginETAL95MonadicNPMonadicCoNP}.  For a presentation
of role analysis and related systems from the perspective of
monadic second-order logic, see
\cite{KuncakRinard02ReasoningHeapHOL}.

In this paper we present a very simple form of constraints
that we call {\em regular graph constraints}.  A set of
regular graph constraints can be specified by a single graph
$G$.  A heap $H$ satisfies the constraints iff there exists
a graph homomorphism from $H$ to $G$.  Regular graph
constraints abstract the problem of mutually recursive
definitions of properties over potentially cyclic graphs.
The existential quantification over predicates is modeled in
regular graph constraints as the existence of a homomorphism
to a given fixed graph.  Regular graph constraints are
closed under conjunction and in certain cases are closed
under disjunction (Section~\ref{sec:closureProperties}).
Moreover, regular graph constraints generalize the notion of
tree
automata~\cite{Thomas97LanguagesAutomataLogic,ComonETAL97Tata}
and domino
systems~\cite{GiammarresiRestivo97TwoDimensionalLanguages},
without going all the way to monadic second-order logic for
richer domains.  In this paper we consider as the domain of
interpretation the class of {\em heaps}.  Our notion of heap
is an abstraction of garbage collected heap in a programming
languages like Java or ML.  Heaps contain a ``root'' node
and a ``null'' node, all nodes are reachable from the root,
and all edges are total functions mapping nodes to nodes.

In Section~\ref{sec:satOverHeaps} we show that there is a
simple and efficient algorithm that decides if regular graph
constraints have a heap model.  This results in a simple
sanity test on regular graph constraint specifications that
rules out the contradictory specifications.

We next turn to the problem of checking if one set of
regular graph constraints implies another set of regular
graph constraints over the set of heaps.  Our main
contribution (Section~\ref{sec:implUndec}) is the proof that
the implication problem is undecidable.  

The implication problem of graphs arises in compositional
checking of programs if procedure preconditions or
postconditions are given as regular graph constraints.  In
Section~\ref{sec:stepwiseInvs} we show that the implication
problem also arises when proving that an invariant holds
after every program step.  These verification problems are
therefore undecidable.  Our result places limitations on the
completeness of systems such as role
analysis~\cite{KuncakETAL02RoleAnalysis} and shape
analysis~\cite{SagivETAL99Parametric} that use homomorphic
images to represent the abstraction of the heap.  The
undecidability of regular graph constraints means that
semantically checking the implication of such homomorphic
graph images is undecidable.

A common way of showing the undecidability of problems over
graphs is to encode the Turing machine computation
histories~\cite{Sipser97TheoryComputation} as a special form
of graphs called {\em grids}.  The difficulty with showing
the undecidability of implication of regular graph
constraints is that regular graph constraints cannot define
the subclass of grids among the class of heaps.  Indeed,
this is why {\em satisfiability} of regular graph
constraints over heaps is {\em decidable}.  To show the {\em
undecidability} of the {\em implication} of regular graph
constraints, we use the constraints on both sides of the
implication to restrict the set of possible counterexample
models for the implication.  For this purpose we introduce a
new class of graphs called {\em corresponder graphs}.
Satisfiability of regular graph constraints over
corresponder graphs mimics the solution of a Post
correspondence problem instance, and is therefore
undecidable.  We give a method for constructing an
implication such that all counterexamples for the validity
of implication are corresponder graphs which satisfy a given
regular graph constraint.  This shows that the validity of
the implication is undecidable.

Due to closure under conjunction, the implication is
reducible to the equivalence of regular graph constraints.
As a result, the equivalence of two regular graph
constraints is also undecidable.


%% file: definitions.tex
\section{Regular Graph Constraints} \label{sec:definitions}

\newcommand{\nullConst}{\m{null}}
\newcommand{\rootConst}{\m{root}}
\newcommand{\implies}{\mbox{ implies }}
\newcommand{\ifandonlyif}{\mbox{ iff }}
\newcommand{\propand}{\mbox{ and }}
\newcommand{\propor}{\mbox{ or }}
\newcommand{\inDegree}{\m{inDegree}}
\newcommand{\partition}{\m{partit}}
\newcommand{\singleton}{\m{singl}}
\newcommand{\homs}{\to}
\newcommand{\Paths}{\m{Paths}}
\newcommand{\word}{\m{word}}
\newcommand{\pin}{\in_p}
\newcommand{\allgraphs}{{\cal G}}
\newcommand{\allheaps}{{\cal H}}
\newcommand{\boolcomb}{{\cal B}}

In this section we define the class of graphs considered in
this paper as well as its subclasses heaps, trees, lists, grids,
and corresponder graphs.  We present our regular graph
constraints, give several equivalent formulations of the
constraints and show that our constraints capture tree
automata and domino systems as special cases.  We then
review some decidability properties, show that
satisfiability of regular constraints over heaps is
efficiently decidable and state some closure properties of
regular graph constraints.

\subsection{Preliminaries}

If $r \subseteq A \times B$ and $S \subseteq A$,
relational image of set $S$ under $r$ is defined as
\[
    r[S] = \{ y \mid x \in S, \tu{x,y} \in r \}
\]
We use $\eop$ to mark the end of a proof and $\eoex$
to mark the end of an example.

\subsection{Graphs}

We will be considering the following class of directed
graphs in this paper.  Our graphs contain two kinds of
edges, which we represent by relations $s_1$ and $s_2$.
These relations may represent fields in an object-oriented
program.  The constant $\rootConst$ represents the root of
the graph.  We use edges terminating at $\nullConst$ to
represent partial functions and abstractions of graphs with
partial functions.
\begin{definition}
A {\em graph} is a relational structure
\[
   G = \tu{V,s_1,s_2,\nullConst,\rootConst}
\]
where
\begin{itemize}
\item $V$ is a finite set of nodes;
\item $\rootConst, \nullConst \in V$ are distinct constants,
      $\rootConst \neq \nullConst$;
\item $s_1, s_2 \subseteq V\times V$ are two kinds of graph edges,
such that for all nodes $x$
\[
   \tu{\nullConst,x} \in s_i  \ifandonlyif  x=\nullConst
\]
for $i \in {1,2}$.
\end{itemize}
We use $\allgraphs$ to denote the class of all graphs.
\end{definition}
An $s_1$-successor of a node $x$ is any element of the set
$s_1[\{x\}]$, similarly an $s_2$-successor of $x$ is any
element of $s_2[\{x\}]$.  Note that there are exactly two
edges originating from $\nullConst$.  When drawing graphs we
never show these two edges.
\begin{definition}
A {\em heap} is a graph $G =
\tu{V,s_1,s_2,\nullConst,\rootConst}$ where relations $s_1$
and $s_2$ are total functions and where for all $x \neq
\nullConst$, node $x$ is reachable from $\rootConst$.
We use $\allheaps$ to denote the class of all heaps.
\end{definition}
\begin{definition}
The in-degree of a node $x$ in a graph is the number of
edges terminating at $x$.
\[
    \inDegree(x) = |\{y \mid \exists i \> \tu{y,x} \in s_i \}|
\]
\end{definition}
\begin{definition}
A {\em tree} is a connected acyclic graph such that
$\inDegree(x) \leq 1$ for every node $x$.
\end{definition}
\begin{definition}
A {\em list} is a tree with at most one non-null outgoing
edge: for every node $x$, $s_1(x)=\nullConst$ or
$s_2(x)=\nullConst$.
\end{definition}

\subsection{Graphs as Constraints}

A regular constraint on a graph $G$ is a constraint stating
that $G$ can be homomorphically mapped to another graph
$G'$.

\begin{definition}
We say that a graph $G$ satisfies the constraints given by a
graph $G'$, and write $G \homs G'$, iff there exists a
homomorphism from $G$ to $G'$.
\end{definition}
A homomorphism between graphs is defined as follows.
\begin{definition} \label{def:homomorphism}
A function $h : V \to V'$ is a homomorphism between
graphs
\[
   G = \tu{V,s_1,s_2,\nullConst,\rootConst}
\]
and
\[
   G' = \tu{V',s'_1,s'_2,\nullConst',\rootConst'}
\]
iff all of the following conditions hold:
\begin{enumerate}
\item[1.] $\tu{x,y} \in s_i  \implies  \tu{h(x),h(y)} \in s'_i$,
      \ for all $i \in \{1,2\}$
\item[2.] $h(x) = \rootConst'$ iff $x = \rootConst$
\item[3.] $h(x) = \nullConst'$ iff $x = \nullConst$
\end{enumerate}
If there exists a homomorphism from $G$ to $G'$, we call
$G$ a {\em model} for $G'$.
\end{definition}
We can think of a homomorphism $h : V \to V'$ as a coloring
of the graph $G$.  The color $h(x)$ of a node $x$ restricts
the colors of the $s_1$-successors of $x$ to the colors in
$s_1[\{h(x)\}]$ and the colors of the $s_2$-successors to the
colors in $s_2[\{h(x)\}]$.
\begin{example}
A graph $G$ can be colored by $k$ colors so that the
adjacent nodes have different colors iff $G$ is homomorphic
to a complete graph without self-loops,
\[
   G' = \tu{V',s'_1,s'_2,\nullConst',\rootConst'}
\]
with $V' = \{1,\ldots,k\}$, and
\[
   s'_1 = s'_2 = \{ \tu{x',y'} \mid x' \neq y' \}.
\]
\end{example}
The identity function is a homomorphism from the graph to
itself.  Therefore, $G \homs G$ for every graph $G$.  The
following fundamental property of homomorphisms also holds.
\begin{proposition}[Homomorphisms compose] \label{prop:homsCompose}
Let
\[\begin{array}{l}
   G = \tu{V_,s_1,s_2,\nullConst,\rootConst} \\
   G' = \tu{V_,s'_1,s'_2,\nullConst',\rootConst'} \\
   G'' = \tu{V_,s''_1,s''_2,\nullConst'',\rootConst''} \\
  \end{array}
\]
and let $h : G \to G'$ and $h' : G' \to G''$ be
homomorphisms.  Then $h_0 : G \to G''$ where $h_0 = h' \circ
h$ is also a homomorphism.
\end{proposition}
A consequence of Proposition~\ref{prop:homsCompose} is that
$\homs$ is a transitive relation.
\begin{definition}[Satisfiability]
A graph $G'$ is satisfiable over the class of graphs $C$ iff
there exists a graph $G \in C$ such that $G \homs G'$. 
Satisfiability problem over the class of graphs $C$ is:
given a graph $G'$, determine if $G'$ is satisfiable.
\end{definition}

\begin{definition}[Implication]
We say that $G_1$ implies $G_2$ over the
class of graphs $C$, and write
\[
    G_1 \graphimplies_C G_2,
\]
iff
\[
    (H \homs G_1) \implies (H \homs G_2)
\]
for all graphs $H \in C$.
\end{definition}
We will omit $C$ in $\graphimplies_C$ if the class of
graphs is clear from the context.

The following fact provides a sufficient condition for the graph
implication to hold.  It is a direct consequence
of Proposition~\ref{prop:homsCompose}.

\begin{proposition} \label{prop:ifHomThenImpl}
Let $C$ be any class of graphs.  Let $G \homs G'$.
Then $G \graphimplies_C G'$.
\end{proposition}
In Section~\ref{sec:implUndec} we show that the implication
of graphs is undecidable over the class of heaps.

\subsection{Paths}

We next state several simple properties of paths that will
be useful in Section~\ref{sec:implUndec}.
\begin{definition}[Path]
Let
\[
    G = \tu{V,s_1,s_2,\nullConst,\rootConst}
\]
be a graph and $n \geq 0$.  A {\em path in graph $G$},
denoted $p \in \Paths(G)$ starting at $x_0$ and terminating
at $x_n$ is a sequence of alternating nodes and labels:
\[
    p = x_0,l_0,x_1,l_1,\ldots,l_{n-1},x_n
\]
such that $x_0,\ldots,x_n \in V$;
$l_0,\ldots,l_{n-1} \in \{1,2\}$ and
$\tu{x_i,x_{i+1}} \in s_i$ for all $i$, $0 \leq i < n$.
We define $\word(p) \in \{1,2\}^{*}$ by
\[
   \word(p) = l_0 l_1 \ldots l_{n-1}
\]
\end{definition}
\begin{definition}[Slice]
A path is a {\em slice} if it starts at $\rootConst$ and
terminates at $\nullConst$.
\end{definition}
\begin{definition}[Path Image]
Let $h$ be a homomorphism from graph $G_0$ to graph $G$ and
let
\[
    p = x_0,l_0,x_1,l_1,\ldots,l_{n-1},x_n
\]
be a path in $G_0$.  Then
\[
   h[p] = h(x_0),l_0,h(x_1),l_1,\ldots,l_{n-1},h(x_n)
\]
is the image of path $p$ under the homomorphism $h$.
\end{definition}
The following facts are a consequence of the definition of
homomorphism.
\begin{proposition} \label{prop:homPaths}
Let $h$ be a homomorphism from graph $G_0$ to graph $G$
and let $p$ be a path in $G_0$.  Then
\begin{enumerate}
\item $h[p]$ is a path in $G$;
\item if $p$ is a slice then $h[p]$ is a slice;
\item $\word(p)=\word(h[p])$.
\end{enumerate}
\end{proposition}
\begin{definition}
Let $p$ be a path in $G$ and $e$ be a regular
expression over the alphabet $\{1,2\}$.  We write
\[
   p \pin e
\]
iff $\word(p)$ belongs to the language of the regular
expression $e$.
\end{definition}
From Proposition~\ref{prop:homPaths} we directly obtain the
following fact.
\begin{proposition}[Regular Expression Test] 
\label{prop:regexpTest}
Let $G_0 \homs G$.  Then if $e$ is any regular
expression over the alphabet $\{1,2\}$ such that
$G_0$ contains some slice $p \pin e$, then $G$ contains
some slice $p' \pin e$.
\end{proposition}
\begin{proof}
Let $p$ be a slice in $G_0$ and $\word(p) \pin e$.  By
Proposition~\ref{prop:homPaths} we have that $h[p]$ is a
slice in $G$ and $\word(h[p]) = \word(p) \pin e$.
\end{proof}

\noindent
We will use the contrapositive of
Proposition~\ref{prop:regexpTest} in the proof of
Proposition~\ref{prop:corrEncoding}
(Section~\ref{sec:implUndec}).

\subsection{Regular Constraints and EMSOL} \label{sec:graphsEMSOL}

We can express the property of being homomorphic to a fixed
graph $G'$ by an existential monadic second-order logic
formula (EMSOL) of a special form.  

Let
\[
   G' = \tu{V',s'_1,s'_2,\nullConst',\rootConst'}
\]
be a fixed graph and let $V' = \{x'_0,\ldots,x'_{k-1}\}$
with $x'_0 = \nullConst'$, $x'_1 = \rootConst'$.
Then~(\ref{eqn:EMSOLencoding}) is a formula in EMSOL
interpreted over the graph
\[
   G = \tu{V,s_1,s_2,\nullConst,\rootConst}
\]
expressing that $G$ is homomorphic to $G'$.  We use the
uppercase identifiers $X_0,\ldots,X_{k-1}$ to denote the
second order variables.  These variables range over the
subsets of $V$.  The lowercase identifiers are first-order
variables ranging over the elements of $V$.  Notation
$X_i(z)$ means that $z$ is an element of the set $X_i$.  The
predicate $s_i(x,y)$ means that $\tu{x,y} \in s_i$ holds in
the graph $G$.  (For the precise definition of monadic
second-order logic see e.g.
\cite{GecsegSteinby97TreeLanguages}, pp28.)
\begin{equation} \label{eqn:EMSOLencoding}
\begin{array}{ll} 
   \exists X_0,\ldots,X_{k-1}. \\ 
    \ \  \partition(X_0,\ldots,X_{k-1}) \land 
         \singleton(X_0,\nullConst) \land 
         \singleton(X_1,\rootConst) \ \land \\
    \ \ \forall x \ \bigwedge\limits_{0 \leq j < k}
          (X_j(x) \Longrightarrow P_j(x)) \\
  \end{array}
\end{equation}
where
\[
   P_j(x) = P^1_j(x) \land P^2_j(x)
\]
\[
   P^i_j(x) = \forall y.\ \ s_i(x,y) \Longrightarrow
     \bigvee_{\begin{array}{c}
              \scriptstyle 0 \leq l < k \\
              \scriptstyle \tu{x'_j,x'_l} \in s'_i
              \end{array}}
        X_l(y)
\]
\[
    \singleton(X,y) = (\forall z.\ X(z) \Longleftrightarrow y=z)
\]
\[\begin{array}{l}
    \partition(Y_0,\ldots,Y_{n-1}) = \mnl
      \forall x.\
            \bigvee\limits_{0 \leq i < n} Y_i(x) \ \land 
            \bigwedge\limits_{0 \leq i < j < n}
             \lnot (Y_i(x) \land Y_j(x))
\end{array}\]

Viewing graph as a formula justifies our previous
definitions of graph satisfiability and implication.  We can
similarly talk about the graph conjunction, disjunction etc.

We may increase the ease of expression of some properties by
relaxing the form of EMSOL formula~(\ref{eqn:EMSOLencoding})
without changing the expressive power.  The reason is that
that the relaxed form can be converted into a normal form
that can be described by a graph homomorphism.  Let
\[
   \boolcomb_f(B_0,B_1,\ldots,B_{n-1};A_0,\ldots,A_{m-1})
\]
denote an arbitrary propositional combination of formulas
$B_0,\ldots,B_{n-1},A_0,\ldots,A_{m-1}$ in which
$B_0,\ldots,B_{n-1}$ occur only negatively.  Then every
formula of the following form is expressible as a graph
constraint.
\begin{equation} \label{eqn:EMSOLflexible}
\begin{array}{ll} 
   \exists X_0,\ldots,X_{k-1}. 
         \singleton(X_0,\nullConst) \land 
         \singleton(X_1,\rootConst) \ \land \mnl
    \ \ \begin{array}{ll}
        \forall x\, \forall y. &
        \boolcomb_1\big( \begin{array}[t]{@{\,}l}
                    s_1(x,y); \\
                    X_0(x),\ldots,X_{k-1}(x), \\
                    X_0(y),\ldots,X_{k-1}(y) \,\big) \ \bigwedge
                    \end{array} \mnl
        & \boolcomb_2\big( \begin{array}[t]{@{\,}l}
                    s_2(x,y); \\
                    X_0(x),\ldots,X_{k-1}(x), \\
                    X_0(y),\ldots,X_{k-1}(y) \,\big)
                    \end{array} \mnl
        \end{array}
  \end{array}
\end{equation}
Compared to~(\ref{eqn:EMSOLencoding}), the
form~(\ref{eqn:EMSOLflexible}) does not require
$X_0,\ldots,X_{k-1}$ to form a partition of the set of all
nodes, it has the quantifiers from $P_j$ lifted to the
topmost level, and allows arbitrary propositional
combinations of predicates $X_i(x)$ and $X_i(y)$. 

\begin{example}
The following formula is of the
form~(\ref{eqn:EMSOLflexible}).  Let us assume that the
formula is interpreted over the class of heaps.  The formula
states that the node $\rootConst$ has in-degree 0.
\[\begin{array}{l}
   \exists X_0,X_1,X_2.\ \ 
         \singleton(X_0,\nullConst) \land 
         \singleton(X_1,\rootConst) \ \land \mnl
    \ \ \begin{array}{ll}
        \forall x\, \forall y.
        \begin{array}[t]{l}
        \lnot (X_1(x) \land X_2(x)) \land 
        (X_0(x) \Rightarrow X_2(x)) \ \land \mnl
        s_1(x,y) \Longrightarrow \big(
         \begin{array}[t]{l}
           (X_1(x) \Rightarrow X_2(y)) \ \land \mnl
           (X_2(x) \Rightarrow X_2(y)) \ \big)
         \end{array} \mnl
        s_2(x,y) \Longrightarrow 
         \begin{array}[t]{l}
           \lnot X_1(y) 
         \end{array}
        \end{array}
        \end{array}
  \end{array}    
\]
The formula uses the set $X_2$ that contains $\nullConst$ as
well as the nodes reachable from $\rootConst$ along the
$s_1$ edges.  The formula states explicitly that $X_1$
and $X_2$ are disjoint.  Because $s_1$ edges from $X_2$ can
only lead to $X_2$, there are no $s_1$ edges to
$\rootConst$.  The constraint that root has no $s_2$ edges
is specified directly, without introducing an auxiliary set
of nodes.  In general, negation and the implicit absence of
constraints are often more convenient to express with a
formula than with a graph homomorphisms.

Note that if we replaced the subformula $(X_0(x) \Rightarrow
X_2(x))$ with $\lnot (X_0(x) \land X_2(x))$ the resulting
formula would require the existence of a cycle in the graph.
\end{example}

\begin{proposition}
The two families of formulas (\ref{eqn:EMSOLencoding}) and
(\ref{eqn:EMSOLflexible}) denote the same family of sets of
graphs.
\end{proposition}
\begin{proof}\ (Sketch)

Given a formula of form (\ref{eqn:EMSOLencoding})
we construct a formula of form (\ref{eqn:EMSOLflexible})
by transforming
\[
    X_j(x) \Rightarrow P^1_j(x) \land P^2_j(x)
\]
into $X_j(x) \Rightarrow P^1_j(x)$ and $X_j(x) \Rightarrow
P^2_j(x)$.  This allows us to write constraints on $s_1$ and
$s_2$ separately.  We then lift the universal quantification
over $y$ to the top level.  The partition constraint is
expressible as a formula that has no occurrences of $s_i$.

Conversely, suppose we are given a formula in form
(\ref{eqn:EMSOLflexible}) and suppose that the formula holds
for a graph $G$ with the set of nodes $V$.  This means that
there exist sets $S_1,\ldots,S_{k-1}$ that satisfy
$\boolcomb_1$ and $\boolcomb_2$.  We construct a family
of sets $T_{i_1,\ldots,i_{k-1}}$ that forms a partition of
set $V$ such that every set $S_j$ is expressible as a
union of some sets $T_{i_1,\ldots,i_{k-1}}$.  Namely, we
define
\[
    T_{i_0,\ldots,i_{k-1}}(y) = 
        S_0^{i_0} \cdots \land \cdots S_{k-1}^{i_{k-1}}
\]
where $i_0,\ldots,i_{k-1} \in \{0,1\}$ and
\[\begin{array}{l}
    S^0 = S \mnl
    S^1 = V \setminus S
\end{array}
\]
This motivates a construction where the second-order
variables $X_0,X_1,X_2,\ldots,X_{k-1}$ are replaced with up
to $2+2^{k-2}$ new variables $X_0,X_1,Y_0,\ldots,Y_{n-1}$.
We then express variables $X_2,\ldots,X_{k-1}$ in terms of
$X_0,X_1,Y_0,\ldots,Y_{n-1}$, write the boolean combinations
$\boolcomb_f$ in disjunctive normal form and use the fact
that $X_0,X_1,Y_0,\ldots,Y_{n-1}$ denote disjoint sets.  As
a result, it is possible to write the original formula in
form~(\ref{eqn:EMSOLencoding}).
\end{proof}

\noindent
Note that, over any class containing all heaps, not every
EMSOL formula corresponds to a regular graph constraint.
This is in contrast with
trees~\cite{Thomas97LanguagesAutomataLogic} and
grids~\cite{GiammarresiRestivo97TwoDimensionalLanguages}.
Even first-order logic can express a constraint that the
graph is a grid.  On the other hand, we have shown in
(\cite{Kuncak01DesigningRoleAnalysis}, pp93) that not even
constraints stronger than regular graph constraints
can express the gridness property.

For a normal form construction using full existential
monadic second-order logic see
see~\cite{SchwentickBarthelmann01LocalNormalForms}.  For
using higher order logic to express heap properties more
general than regular graph constraints, see
\cite{KuncakRinard02ReasoningHeapHOL}.

\subsection{Related Systems}

In this section we show the relationship of our regular
graph constraints with some other systems for defining sets
of graphs.  We also illustrate that decidability of
satisfiability and implication are sensitive to the
subclass of the graphs considered, and change in a
non-monotonic way.

\subsubsection{Words}

Regular graph constraints over lists correspond to regular
word languages.  A regular graph constraint corresponds
to a nondeterministic finite state automaton with the
initial state $\rootConst$ and the final state $\nullConst$.

\subsubsection{Trees}

Satisfiability and implication of regular graph
constraints are decidable over the class of trees.  The
reason is that the entire MSOL is decidable over
trees~\cite{Thomas97LanguagesAutomataLogic}, and regular
graph constraints are expressible in MSOL.

\subsubsection{Pictures}

Domino systems
\cite{GiammarresiRestivo97TwoDimensionalLanguages} are
regular graph constraints over the grids.
\begin{definition}
A grid $m \times n$ is a graph isomorphic to
\[
    G = \tu{V,s_1,s_2,\nullConst,\rootConst}
\]
where
\begin{eqnarray*}
   V &=& \{1,\ldots,m\} \times \{1,\ldots,n\} \\
   s_1 &=& \{ \tu{\tu{i,j},\tu{i,j+1}} \mid
            1 \leq i \leq m; 1 \leq j \leq n-1 \} \ \cup \\
       & & \{ \tu{\tu{i,n},\nullConst} \mid 1 \leq i \leq m \} \\
   s_2 &=& \{ \tu{\tu{i,j},\tu{i+j,j}} \mid 
            1 \leq i \leq m-1; 1 \leq j \leq n  \} \ \cup \\
       & & \{ \tu{\tu{j,m},\nullConst} \mid 1 \leq j \leq n \} \\
   \rootConst &=& \tu{1,1} \\
\end{eqnarray*}

\end{definition}

The chapter
\cite{GiammarresiRestivo97TwoDimensionalLanguages} uses the term
{\em pictures} for grids.  It is easy to see that over the
domain of grids, regular graph constraint are
equivalent to a domino system with $s_1$ edges denoting
horizontal dominoes and $s_2$ edges denoting vertical
dominoes.  The graph homomorphism corresponds to the use of
projection.

\cite{GiammarresiRestivo97TwoDimensionalLanguages} states the
equivalence of domino systems over pictures with
negation-free regular expressions with projections, on-line
tessellation automata, existential monadic second-order
formulas and tiling systems.

We view the fact that, over the grids, regular graph
constraints are equivalent to each of the systems above as
an indication that the definition of regular graph
constraints is natural.

\paragraph{Note} When comparing our regular graph
constraints to trees and domino systems, we notice that in
our definition of a model there are no fixed labels
associated with nodes.  The only labeling of nodes comes
from the graph homomorphism, which corresponds to projection
in tree and picture languages.  Our simplification makes our
undecidability result strictly stronger.  Furthermore,
regular graph constraints can capture the distinction
between a node with an edge terminating at $\nullConst$ and
a node with an edge terminating at a node that is not null.
This distinction can be used for encoding in the structure
of the graph any fixed labeling of graph nodes.

\subsection{Decidability of Implication over Graphs}

Satisfiability problem over the class of graphs is
trivial. Namely, $G \homs G$, so every graph is satisfiable.
The implication problem over graphs is also decidable, in
contrast to the implication problem over the class of heaps,
which we will show undecidable in
Section~\ref{sec:implUndec}.

\begin{proposition}
\[
    G_1 \graphimplies_{\allgraphs} G_2  \ifandonlyif G_1 \homs G_2
\]
\end{proposition}
\begin{proof}
Let $G_1 \graphimplies_{\allgraphs} G_2$.  Because $G_1 \homs G_1$,
we obtain $G_1 \homs G_2$.  Conversely, let $G_1 \homs G_2$.
By Proposition~\ref{prop:ifHomThenImpl} we conclude
$G_1 \graphimplies_{\allgraphs} G_2$.
\end{proof}

\noindent
Our regular graph constraints are weaker than finite graph
acceptors of \cite{Thomas91LogicsTilingsAutomata}.  It is
easy to see that finite graph acceptors can define the
gridness property.  Therefore, domino system satisfiability
is reducible to satisfiability of finite graph acceptors,
which makes finite graph acceptor satisfiability
undecidable.

\subsection{Satisfiability over Heaps} \label{sec:satOverHeaps}

We show that satisfiability for heaps is efficiently
decidable by the nondeterministic algorithm in
Figure~\ref{fig:heapSatCheck}.  The goal of the algorithm is
to find, given a graph $G'$, whether there exists a heap $G$
such that $G \homs G'$.  Recall that the property of a heap
is that every node has exactly one $s_1$ outgoing edge and
exactly one $s_2$ outgoing edge.  This property need not be
satisfied by $G'$, so we cannot take $G=G'$ to be the heap
proving satisfiability of $G'$.  The algorithm updates
the current graph until it becomes a heap or an empty graph.
(For the purpose of this algorithm we allow even
$\nullConst$ and $\rootConst$ to be removed from the graph.)
If nonempty, the result is a heap $G$ such that $G \homs
G'$.

\begin{figure}
{\sf
{\bf GraphCleanup:}
Repeat the following operations until the
graph stabilizes:
\begin{enumerate}
\item remove an unreachable node
\item remove a node $x$ such that $s_1[\{x\}]=\emptyset$
or $s_2[\{x\}]=\emptyset$
\end{enumerate}
{\bf mark(x):}
\begin{enumerate}
\item if $x$ is marked then return, otherwise:
\item $\m{select}(x)$;
\item pick a $s_1$-successor $y$ of $x$; $\m{select}(\tu{x,y})$; 
$\m{mark}(y)$
\item pick a $s_2$-successor $z$ of $x$; $\m{select}(\tu{x,y})$; 
$\m{mark}(z)$
\end{enumerate}
{\bf SatisfiabilityCheck:}
Repeat the following operations until the graph stabilizes:
\begin{enumerate}
\item perform GraphCleanup;
\item if the resulting graph is empty, then $G'$ is unsatisfiable;
\item otherwise a heap satisfying $G'$ can be obtained as follows:
\item let all graph nodes be unmarked;
\item $\m{mark}(\rootConst)$;
\item return subgraph containing selected nodes
\end{enumerate}
}
\caption{Satisfiability check for Heaps\label{fig:heapSatCheck}}
\end{figure}

\begin{proposition}  \label{prop:heapSatOk}
The procedure in Figure~\ref{fig:heapSatCheck} is a correct
algorithm for determining satisfiability of a graph over the
class of heaps.
\end{proposition}
\begin{proof}
The procedure consists of two parts: GraphCleanup and
SatisfiabilityCheck.  The GraphCleanup part eliminates
useless nodes and determines whether there exists a heap $H$
such that $H \homs G'$.  Graph Cleanup terminates because it
decreases the size of the current graph $G$ in every step.
The \m{mark} phase terminates because it does a simple
breadth-first search.

Observe that GraphCleanup does not reduce the set of heaps
homomorphic to $G'$.  Namely, if a node $x$ of $G$ is
removed in GraphCleanup, then no node is mapped to $x$
under any homomorphism $h$.  Therefore, if GraphCleanup
returns an empty graph, then $G'$ is unsatisfiable.

Assume that GraphCleanup returns a nonempty graph $G$.
Then $G$ contains root and every node in $x$ has a
$s_1$-successor and a $s_2$-successor, but some of the nodes may
have two $s_1$-successors or $s_2$-successors.  Invoking $\m{mark}$
will do a depth-first search on $G$ and pick a subgraph $H$
where every node has exactly one $s_1$-successor and one
$s_2$-successor.  The resulting graph $H$ will therefore be a
heap.  We have $H
\homs G'$ because $H$ is a subgraph of $G'$.
\end{proof}

\subsection{Closure Properties} \label{sec:closureProperties}

In this section we give a construction for computing the
conjunction of two graphs and a construction for computing
the disjunction of two graphs.  We will use these
constructions in Section~\ref{sec:implUndec}.

\subsubsection{Conjunction}

We show how to use a Cartesian product construction to
obtain a conjunction of two graphs $G_1$ and $G_2$.

\begin{definition}[Cartesian Product] \label{def:product}
Let
\[\begin{array}{l}
   G^1 = \tu{V^1,s^1_1,s^1_2,\nullConst^1,\rootConst^1} \\
   G^2 = \tu{V^2,s^2_1,s^2_2,\nullConst^2,\rootConst^2} \\
  \end{array}
\]
be graphs.  Then $G^0 = G^1 \times G^2$ is the graph
\[
   G^0 = \tu{V^0,s^0_1,s^0_2,\nullConst^0,\rootConst^0} \\
\]
such that:
\[\begin{array}{rcl}
    \nullConst^0 &=& \tu{\nullConst^1,\nullConst^2} \\[\glngt]
    \rootConst^0 &=& \tu{\rootConst^1,\rootConst^2} \\[\glngt]
    V^0 &=& \{\nullConst^0,\rootConst^0\} \\[\glngts]
    &\cup& (V^1 \setminus \{\nullConst^1,\rootConst^1\}) \times
           (V^2 \setminus \{\nullConst^2,\rootConst^2\}) \\[\glngt]
    s_i^0 &=& \{ \tu{\tu{x^1,x^2},\tu{y^1,y^2}} \mid
               \tu{x^1,y^1} \in s_i^1;
               \tu{x^2,y^2} \in s_i^2 \}, \\[\glngts]
     & & i \in \{ 1,2 \} \\[\glngt]
  \end{array}
\]
\end{definition}

\begin{proposition}[Conjunction via Product] 
\label{prop:conjunction}
For every graph $G$,
\[
    G \homs G_1 \times G_2\ \ \ifandonlyif \ \
        G \homs G_1 \propand G \homs G_2
\]
In other words, $G_1 \times G_2$ is a conjunction of
$G_1$ and $G_2$.
\end{proposition}
\begin{proof}
$(\Longrightarrow):$  Let $G \homs G^1$ and $G \homs G^2$
with $h^1 : V \to V^1$ and $h^2 : V \to V^2$ where $G^1$
and $G^2$ are defined as in Definition~\ref{def:product}
and
\[
   G = \tu{V,s_1,s_2,\nullConst,\rootConst}
\]
Let $h^0 = h^1 \times h^2$ where
\[
   (h^1 \times h^2)(x) = \tu{h^1(x),h^2(x)}
\]
We claim that $h^0$ is a homomorphism from $G$ to $G^0 = G^1
\times G^2$.  It is straightforward to verify that properties
2 and 3 of homomorphism hold for $h^0$.  Let $i \in \{1,2\}$.
If $\tu{x,y} \in s_i$ then $\tu{h^1(x),h^1(y)}
\in s^1_i$ and $\tu{h^2(x),h^2(y)} \in s^2_i$.
Because $h^1$ and $h^2$ satisfy properties 2 and 3 of
homomorphism, we have $\tu{h^1(x),h^2(x)},
\tu{h^2(y),h^2(y)} \in V^0$ regardless of whether $x, y \in
\{\nullConst,\rootConst\}$ or not.  
We therefore conclude
$\tu{\tu{h^1(x),h^2(x)},\tu{h^1(x),h^2(y)}} \in s^0_i$ by
the definition of $s^0_i$.  Hence, $h^0$ is a homomorphism and $G
\homs G^0$.  

$(\Longleftarrow):$ Let $G \homs G^0$ with $h^0 : V \to
V^0$.  Define $h^1 = \pi_1 \circ h$ and $h^2 = \pi_2 \circ
h$.  Let's show that $h^1$ is a homomorphism (an analogous
argument holds for $h_2$).  It is straightforward
to see that properties 2 and 3 hold for $h_1$.  For property 1,
let $i \in \{1,2\}$ and let $\tu{x,y} \in s_i$.  Let
$h^0(x) = \tu{x^1,x^2}$ and $h^0(y) = \tu{y^1,y^2}$.
Because $h^0$ is a homomorphism, we have
\[
    \tu{\tu{x^1,x^2}, \tu{y^1,y^2}} \in s^0_i
\]
By definition of $s^0_i$ we conclude
$\tu{x^1,y^1} \in s^1_i$ which by definition of $h^1$ means
$\tu{h^1(x),h^1(y)} \in s^1_i$.
\end{proof}

\subsubsection{Disjunction}

Given our definition of graphs, there is no construction
that would yield disjunction of arbitrary graphs over the
family that contains all heaps.  We illustrate this fact
with an example.  We then give a simple condition on graphs
that ensures that the disjunction construction is possible
over the domain of heaps.

\begin{example}
Let
\[\begin{array}{rcl}
    G^1 &=& \tu{\{\rootConst,\nullConst\}, s_1^1, s_2^2,
                \rootConst, \nullConst} \\
    s_1^1 &=& \{ \tu{\rootConst,\nullConst} \} \\
    s_2^1 &=& \{ \tu{\rootConst,\rootConst} \} \\
\end{array}\]
and
\[\begin{array}{rcl}
    G^2 &=& \tu{\{\rootConst,\nullConst\}, s_1^2, s_2^2,
                \rootConst, \nullConst} \\
    s_1^2 &=& \{ \tu{\rootConst,\rootConst} \} \\
    s_2^2 &=& \{ \tu{\rootConst,\nullConst} \} \\
\end{array}\]
In the class of heaps, the only model for $G^1$ is $G^1$
itself, and the only model of $G^2$ is $G^2$.  Now assume
that there exists graph
\[ 
   G^0 = \tu{V^0,s_1^0,s_2^0,\nullConst,\rootConst}
\]
such that $G^1 \homs G^0$ and
$G^2 \homs G^0$.  From $G^1 \homs G^0$ we conclude
\[
    \tu{\rootConst,\nullConst} \in s_1^0
\]
and from $G^2 \homs G^0$ we conclude
\[
    \tu{\rootConst,\nullConst} \in s_2^0
\]
Therefore for the graph
\[\begin{array}{rcl}
    G^3  &=& \tu{\{\rootConst,\nullConst\}, s_1^3, s_2^3,
                \rootConst, \nullConst} \\
    s_1^3 &=& \{ \tu{\rootConst,\nullConst} \} \\
    s_2^3 &=& \{ \tu{\rootConst,\nullConst} \} \\
\end{array}\]
we have $G_3 \homs G^0$ as well.  So there is no
graph $G^0$ such that for all heaps $G$,
\[
    (G \homs G^0) \ \ifandonlyif \ 
     ((G \homs G^1) \propor (G \homs G^2))
\]
\end{example}
To ensure that we can find union graphs over the set of
heaps, we will require $s_2(\rootConst)=\nullConst$.

\begin{definition}[Orable Graphs]
A graph
\[
    G = \tu{V,s_1,s_2,\nullConst,\rootConst}
\]
is {\em orable} iff for all $x \in V$,
\[
    \tu{\rootConst,x} \in s_2 \ \ifandonlyif \ x=\nullConst
\]
\end{definition}

\begin{figure}
\begin{center}
\includegraphics[scale=0.25]{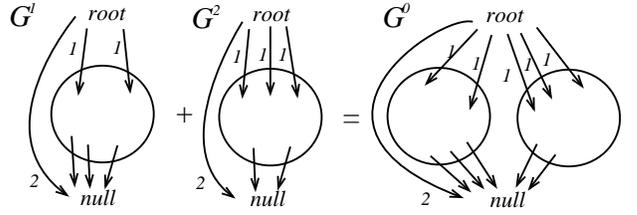}
\end{center}
\caption{Graph Sum}
\end{figure}

\begin{definition}[Graph Sum] \label{def:graphSum}
Let
\[\begin{array}{l}
   G^1 = \tu{V^1,s^1_1,s^1_2,\nullConst,\rootConst} \\
   G^2 = \tu{V^2,s^2_1,s^2_2,\nullConst,\rootConst} \\
  \end{array}
\]
be orable graphs such that $V^1 \cap V^2 =
\{\nullConst,\rootConst\}$.  Then $G^0 = G^1 + G^2$ is the
graph
\[
   G^0 = \tu{V^0,s^0_1,s^0_2,\nullConst,\rootConst} \\
\]
where
\[\begin{array}{rcl}
  V^0 &=& V^1 \cup V^2 \\
  s_1^0 &=& s_1^1 \cup s_2^2 \\
  s_2^0 &=& s_2^1 \cup s_2^2 \\
\end{array}\]
\end{definition}
The following simple fact allows us to form arbitrary finite
sums of orable graphs.
\begin{proposition}
If $G^1$ and $G^2$ are orable graphs, then $G^1+G^2$ is also
orable.
\end{proposition}

\begin{proposition}[Disjunction via Sum]\allowspace
Let $G$ be a heap and $G^1$ and $G^2$ be orable graphs.
Then
\[
   G \homs G^1 + G^2 \ \ \ifandonlyif \ \ 
   G \homs G^1 \propor G \homs G^2
\]
\end{proposition}
\begin{proof}
$(\Longleftarrow):$ Assume without loss of generality $G
\homs G^1$.  Because $G^1$ is a subgraph of $G^1+G^2$, there
exists an identity homomorphism from $G^1$ into $G^1+G^2$.
By Proposition~\ref{prop:homsCompose} we conclude $G \homs
G^1 + G^2$.

$(\Longrightarrow):$
Let $G^0 = G^1 + G^2$ where $G^1$ and $G^2$ are as
as in Definition~\ref{def:graphSum},
and $G \homs G^0$ with a homomorphism
$h : V \to V^0$ where
\[
   G = \tu{V,s_1,s_2,\nullConst,\rootConst}
\]
We claim
\begin{equation} \label{eqn:homChooses}
   h[V] \subseteq V^1  \ \propor \  h[V] \subseteq V^2
\end{equation}
Suppose the claim does not hold.  Then there exist
$x^0, y^0 \in V$ such that
\[\begin{array}{rcl}
    x^1 &\in& V^1 \setminus V^2  \\
    y^2 &\in& V^2 \setminus V^1  \\   
\end{array}\]
where $x^1 = h(x^0)$ and $y^2 = h(y^0)$.  By definition of
heap, there exists a sequence of nodes $p =
\rootConst,z^0,\ldots,x^0$ forming a path from $\rootConst$ to
$x^0$ in $G$.  Because $x^0 \notin
\{\nullConst,\rootConst\}$, the path has length at least
two and $z^0 \notin \{\nullConst,\rootConst\}$ (it may or
may not be $z^0=x^0$).  Because $G^1+G^2$ is orable, the
edge from $\rootConst$ to $z^0$ cannot be from $s_2$, so
$s_1(\rootConst) = z^0$.  We claim $h(z^0) \in V^1 \setminus
V^2$.  Indeed, suppose $h(z^0) \in V^2$.  By the properties
of homomorphism and because in $G^1+G^2$ there are no edges
from nodes $V^1 \setminus \{\nullConst,\rootConst\}$ to
nodes $V^2 \setminus \{\nullConst,\rootConst\}$, we have
$h(w) \in V^2$ for every node $w$ of the path $p$.  This is
a contradiction with $h(x^0) \in V^1 \setminus V^2$.  We
conclude
\[
   z^0 \in V^1 \setminus V^2
\]
Repeating an analogous argument for node $y^0$, we conclude
\[
   z^0 \in V^2 \setminus V^1
\]
because $z^0 = s_1(\rootConst)$ is the unique $s_1$-successor of
$\rootConst$ in the heap $G$.  We have arrived at
contradiction, so (\ref{eqn:homChooses}) is true.  If $h[V]
\subseteq V^1$ then $G \homs G^1$ and if $h[V] \subseteq
V^2$ then $G \homs G^2$.
\end{proof}

A product of orable graphs is orable.  
\begin{proposition}
Let $G^1$ and $G^2$ be orable graphs.  Then $G^1 \times G^2$
is orable.
\end{proposition}
In the sequel we will deal only with orable graphs.



%% file: implication.tex
\section{Undecidability of Implication} \label{sec:implUndec}

\newcommand{\CG}{\m{CG}}
\newcommand{\negno}{16}
\newcommand{\mypic}[1]{\includegraphics[scale=0.25]{#1}}

This section presents the central result of this paper: {\em
The implication of graphs is undecidable over the class of
heaps.}  Our proof proceeds in two steps.  We first
introduce a family of {\em corresponder graphs}.  We show
that satisfiability of graphs over the family of
corresponder graphs is undecidable.

In the second step we show that satisfiability over
corresponder graphs can be reduced the question of whether an
implication between two graphs fails to hold.  The key to
the construction in the second step is that a conjunction of
certain regular graph constraints and negations of regular
graph constraints can precisely characterize the class of
corresponder graphs.

\subsection{Corresponder Graphs}

\begin{figure*}
\begin{center}
\mypic{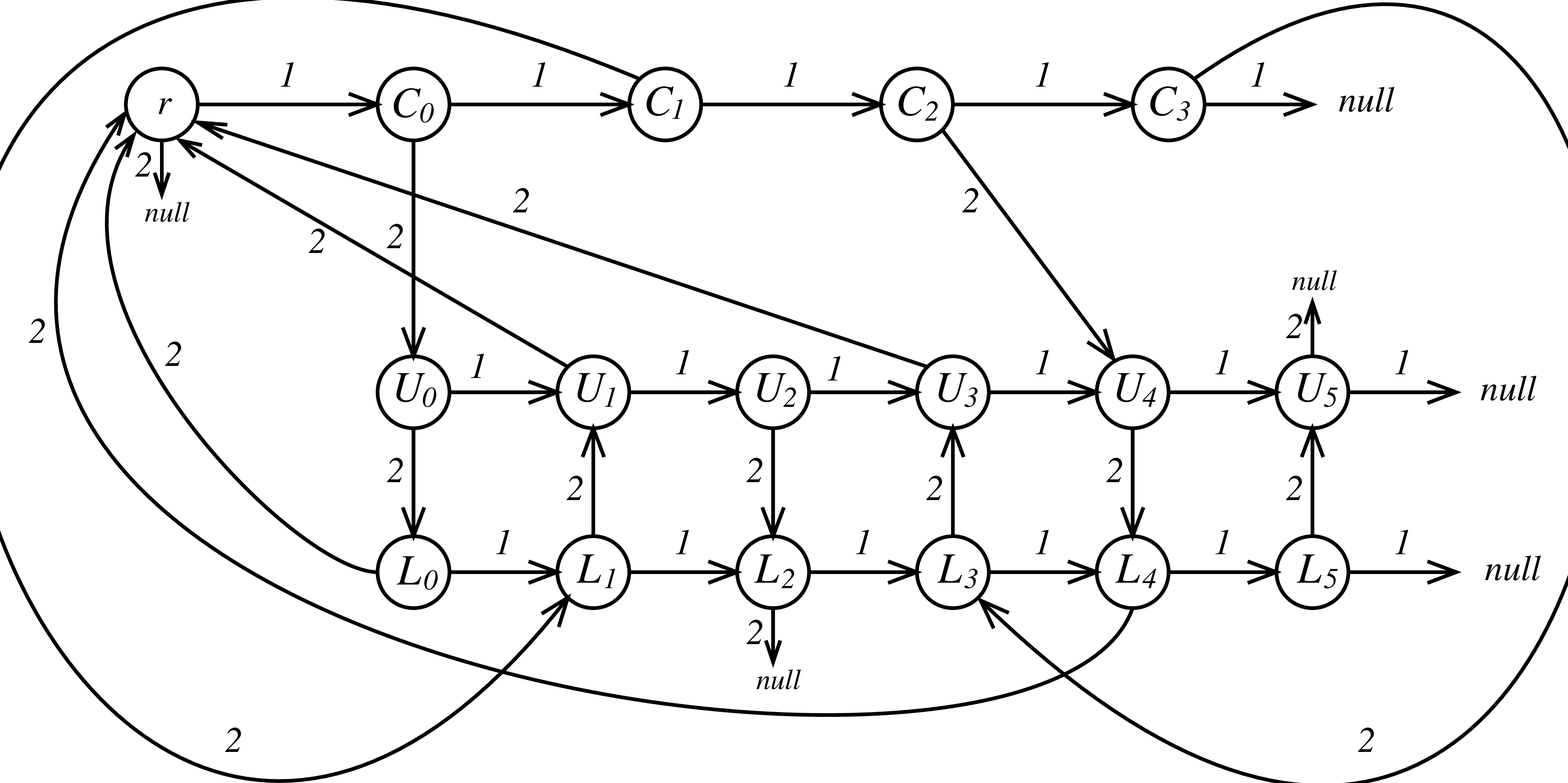}
\end{center}
\caption{An Example Corresponder Graph ($k=2$, $n=3$)
\label{fig:corresponderExample}}
\end{figure*}

Corresponder graphs are a subclass of the class of heaps.
Figure~\ref{fig:corresponderExample} shows an example
corresponder graph.

\begin{definition}
Let $k \geq 2$, $n \geq 2$, and
\[\begin{array}{rcccccccccl}
0 &=& u_0 & < & u_1 & < & \ldots & < & u_{k-1} & < & n \\
0 &=& l_0 & < & l_1 & < & \ldots & < & l_{k-1} & < & n \\
\end{array}\]
A corresponder graph
\[
    \CG(n,k,u_1,\ldots,u_{k-1},l_1,\ldots,l_{k-1})
\]
is a graph isomorphic to
\[
    G = \tu{V,s_1,s_2,\nullConst,\rootConst}
\]
where
\begin{eqnarray*}
V &=& \{ \nullConst, \rootConst \} \\
   &\cup& \{ C_0,C_1,\ldots,C_{2k-1} \} \\
   &\cup& \{ U_0,U_1,\ldots,U_{2n-1} \} \\
   &\cup& \{ L_0,L_1,\ldots,L_{2n-1} \} \\
s_1 &=& \{ \tu{\rootConst, C_0} \} \\
    &\cup& \{ \tu{C_i,C_{i+1}} \mid 0 \leq i < 2k-1 \} \\
    &\cup& \{ \tu{C_{2k-1},\nullConst} \} \\
    &\cup& \{ \tu{U_i,U_{i+1}} \mid 0 \leq i < 2n-1 \} \\
    &\cup& \{ \tu{U_{2n-1},\nullConst} \} \\
    &\cup& \{ \tu{L_i,L_{i+1}} \mid 0 \leq i < 2n-1 \} \\
    &\cup& \{ \tu{L_{2n-1},\nullConst} \} \\
s_2 &=& \{ \tu{\rootConst,\nullConst} \} \\
    &\cup& \{ \tu{C_{2i},U_{2u_i}} \mid 0 \leq i < k \} \\
    &\cup& \{ \tu{C_{2i+1},L_{2l_i+1}} \mid 0 \leq i < k \} \\
    &\cup& \{ \tu{U_{2i}, L_{2i}} \mid 0 \leq i < n \} \\
    &\cup& \{ \tu{L_{2i+1},U_{2i+1}} \mid 0 \leq i < n \} \\
    &\cup& \big\{ \tu{U_{2i+1},\nullConst} \mid 
              i \in \{ 0, \ldots, n-1 \} \setminus
                    \{ l_0,\ldots,l_{k-1} \} \big\} \\
    &\cup& \big\{ \tu{U_{2i+1},\rootConst} \mid 
              i \in \{ l_0,\ldots,l_{k-1} \} \big\} \\
    &\cup& \{ \tu{L_{2i},\nullConst} \mid 
              i \in \{ 0, \ldots, n-1 \} \setminus
                    \{ u_0,\ldots,u_{k-1} \} \big\} \\
    &\cup& \big\{ \tu{L_{2i},\rootConst} \mid 
              i \in \{ u_0,\ldots,u_{k-1} \} \big\} \\
\end{eqnarray*}
The family $\CG$ is the union of all corresponder graphs
$\CG(n,k,u_1,\ldots,u_{k-1},l_1,\ldots,l_{k-1})$.
\end{definition}

\subsection{Corresponder Graph Satisfiability}

For completeness we define the Post correspondence problem,
PCP (\cite{Sipser97TheoryComputation}, pp183).

\begin{definition}
A PCP {\em instance} is a sequence of pairs of nonempty
words:
\[
    \tu{v_0,w_0},\tu{v_1,w_1},\ldots,\tu{v_{m-1},w_{m-1}}
\]
A {\em solution} for a PCP instance is a sequence
\[
    t_0,t_1,\ldots,t_{k-1}
\]
such that
\[
    v_{t_0} v_{t_1} \ldots v_{t_{k-1}} =
    w_{t_0} w_{t_1} \ldots w_{t_{k-1}}
\]
\end{definition}
\cite{Sipser97TheoryComputation} contains the proof for the
following theorem.
\begin{theorem}
The following problem is undecidable: given a PCP instance,
does it have a solution.
\end{theorem}
We will use the following proposition to establish
undecidability of graph implication.
\begin{proposition} \label{prop:PCPreduction}
Satisfiability of graphs over the class of corresponder
graphs is undecidable.
\end{proposition}
\begin{proof}
We give a reduction from PCP.
Let $m \geq 2$
and let
\[
    \tu{v_0,w_0},\tu{v_1,w_1},\ldots,\tu{v_{m-1},w_{m-1}}
\]
be an instance of PCP where $v_i$, $w_i$ are nonempty words.
Introduce names $a_i^j$ and $b_i^j$ for letters in words
$v_i, w_i$:
\[\begin{array}{rcll}
v_i &=& v_i^0 v_i^1 \ldots v_i^{p_i-1} & 0 \leq i \leq m-1 \\
w_i &=& w_i^0 w_i^1 \ldots w_i^{q_i-1} & 0 \leq i \leq m-1 \\
\end{array}\]
where $p_i = |v_i|$ and $q_i = |w_i|$.  We construct a graph
$G$ such that there exists a corresponder graph $G_0$ with
the property $G_0 \homs G$ iff the PCP instance has a
solution.  

\begin{figure*}
\begin{center}
\mypic{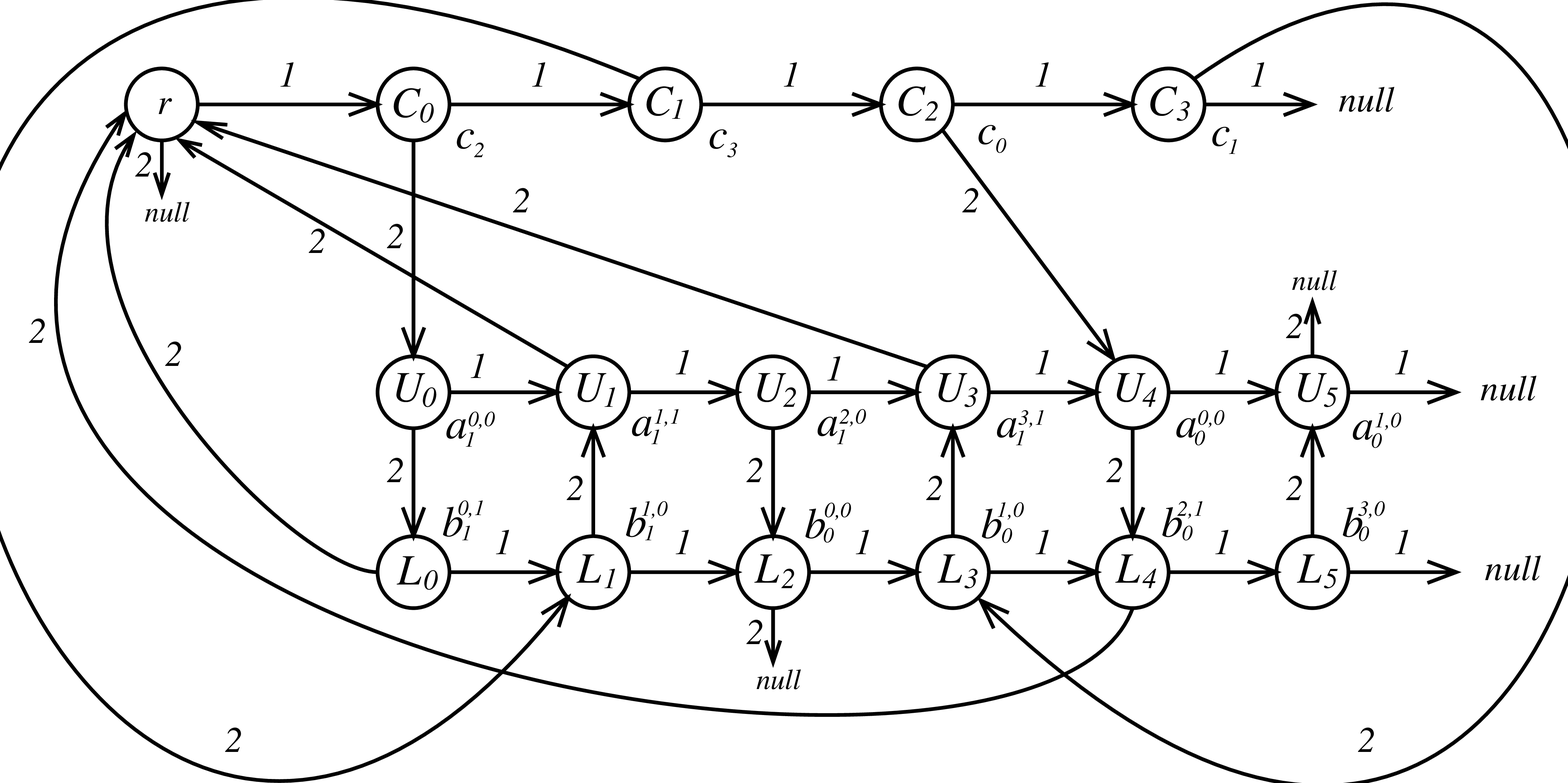}
\end{center}
\caption{Corresponder Graph with a Homomorphism Encoding a Solution of $\tu{c,bc},\tu{ab,a}$
\label{fig:corresponderWithHom}}
\end{figure*}

Figure~\ref{fig:corresponderWithHom} illustrates how a
corresponder graph $G_0$ with a homomorphism from $G_0$ to
$G$ encodes a solution of the PCP instance
$\tu{c,bc},\tu{ab,a}$.

Let
\[
    G = \tu{V,s_1,s_2,\nullConst,\rootConst}
\]
Define the components of $G$ are as follows.  For every pair
of words $\tu{v_i,w_i}$ introduce two nodes $c_{2i},
c_{2i+1} \in V$.  These nodes will summarize $C$-nodes of a
corresponder graph.  For every position $v_i^j$ of the word
$v_i$ introduce nodes $a_i^{2j,0}$ and $a_i^{2j+1,0}$ and
for every position $w_i^j$ introduce nodes $b_i^{2j,0}$ and
$b_i^{2j+1,0}$.  The $a$-nodes will summarize $U$-nodes and
the $b$-nodes will summarize the $L$-nodes of the
corresponder graph.  Introduce also the additional nodes
$b_i^{2j,1}$ to encode the information that $a_i^{2j,0}$
node has an incoming edge from a $c$-node.  As we will see
below, the $b_i^{2j,1}$ nodes have $s_2$ pointing to
$\rootConst$ as opposed to $\nullConst$, which ensures that
every $a_i^{0,0}$ node has an incoming edge from a $c$-node.
For analogous reasons we introduce $a_i^{2j+1,1}$ nodes.
Let
\begin{eqnarray*}
V & =  & \{ \nullConst, \rootConst \} \\
  &\cup& \{ c_0,c_1,\ldots,c_{2m-1} \} \\
  &\cup& \{ a_i^{j,0} \mid 0 \leq i < m; 0 \leq j < 2p_i \} \\
  &\cup& \{ b_i^{j,0} \mid 0 \leq i < m; 0 \leq j < 2q_i \} \\
& \cup & \{ b_i^{2j,1} \mid 0 \leq i < m; 0 \leq j < q_i \} \\
& \cup & \{ a_i^{2j+1,1} \mid 0 \leq i < m; 0 \leq j < p_i \} \\
\end{eqnarray*}
Define $s_1$ graph edges as follows.  

The $c_i$ nodes are connected into a list that begins with
$\rootConst$ and every $c_{2i}$ is followed by $c_{2i+1}$.
The pairs $c_{2i},c_{2i+1}$ for different $i$ can repeat in
the list any number of times and in arbitrary order.  This
list will encode a PCP instance solution.

The nodes representing word positions are linked in the
order in which they appear in the word.  The last position
in a word can be followed by the first position of any other
word, or by $\nullConst$.  The nodes for the $v_i$ words and
the nodes for the $w_i$ words form disjoint lists along the
$s_1$ edges.  
{\newcommand{\lngt}{1.5ex}
\[\begin{array}{r@{\,}c@{\,}l}
s_1 & = & \{ \tu{\rootConst,c_{2i}} \mid 0 \leq i < m \} \\[\lngt]
& \cup  & \{ \tu{c_{2i},c_{2i+1}} \mid 0 \leq i < m \} \\[\lngt]
& \cup  & \{ \tu{c_{2i+1},c_{2j}} \mid 0 \leq i,j < m \} \\[\lngt]
& \cup  & \{ \tu{c_{2i+1},\nullConst} \mid 0 \leq i < m \} \\[\lngt]
& \cup  & \{ \tu{a_i^{2j,0},a_i^{2j+1,\alpha}} \mid 
   0 \leq i < m; 0 \leq j < p_i; \alpha \in \{0,1\} \} \\[\lngt]
& \cup  & \{ \tu{a_i^{2j+1,\alpha},a_i^{2j+2,0}} \mid 
   0 \leq i < m; 0 \leq j < p_i-1; \\[\lngt]
&& \qquad\qqquad \alpha \in \{0,1\} \} \\[\lngt]
& \cup  & \{ \tu{a_i^{2p_i-1,\alpha},a_j^{0,0}} \mid 
   0 \leq i,j < m; \alpha \in \{0,1\} \} \\[\lngt]
& \cup  & \{ \tu{a_i^{2p_i-1,\alpha},\nullConst} \mid 0 \leq i < m ; \alpha \in \{0,1\} \} \\[\lngt]
& \cup  & \{ \tu{b_i^{2j,\alpha},b_i^{2j+1,0}} \mid 
   0 \leq i < m; 0 \leq j < q_i; \alpha \in \{0,1\} \} \\[\lngt]
& \cup  & \{ \tu{b_i^{2j+1,0},b_i^{2j+2,\alpha}} \mid 
   0 \leq i < m; 0 \leq j < q_i-1; \\[\lngt]
&& \qquad\qqquad \alpha \in \{0,1\} \} \\[\lngt]
& \cup  & \{ \tu{b_i^{2q_i-1,0},b_j^{0,\alpha}} \mid 
   0 \leq i,j < m; \alpha \in \{0,1\} \} \\[\lngt]
& \cup  & \{ \tu{b_i^{2q_i-1,0},\nullConst} \mid 0 \leq i < m \} \\[\lngt]
\end{array}\]
}
We define $s_2$ graph edges as follows.

Every $c_j$ edge points to the position at the beginning of
the word.  Even numbered nodes point to the $a^0$-positions;
odd numbered nodes point to $b^1$-positions.

The $a_i$ and $b_j$ word positions are connected so that an
$a$-node points to a $b$-node for even indices, whereas a
$b$-node points to an $a$-node for odd indices.
{\newcommand{\lngt}{1.5ex}
\[\begin{array}{r@{\,}c@{\,}l}
s_2 & = & \{ \tu{\rootConst,\nullConst} \} \\[\lngt]
& \cup  & \{ \tu{c_{2i},a_i^{0,0}} \mid 0 \leq i < m \} \\[\lngt]
& \cup  & \{ \tu{c_{2i+1},b_i^{1,0}} \mid 0 \leq i < m \} \\[\lngt]
& \cup  & \{ \tu{a_i^{0,0},b_k^{2l,1}} \mid 0 \leq i,k < m;
              0 \leq l < q_k; v_i^0 = w_k^l \} \\[\lngt]
& \cup  & \{ \tu{a_i^{2j,0},b_k^{2l,0}} \mid 0 \leq i,k < m;
             0 < j < p_i; 0 \leq l < q_k; \\[\lngt]
&       & \quad\qquad\qquad v_i^j = w_k^l \} \\[\lngt]
& \cup  & \{ \tu{b_k^{2l,0},\nullConst} \mid 0 \leq k < m;
             0 \leq l < q_k \} \\[\lngt]
& \cup  & \{ \tu{b_k^{2l,1},\rootConst} \mid 0 \leq k < m;
             0 \leq l < q_k \} \\[\lngt]
& \cup  & \{ \tu{b_k^{1,0},a_i^{2j+1,1}} \mid 0 \leq i,k < m;
             0 \leq j < p_i; v_i^j = w_k^l \} \\[\lngt]
& \cup  & \{ \tu{b_k^{2l+1,0},a_i^{2j+1,0}} \mid 0 \leq i,k < m;
             0 \leq j < p_i;  \\[\lngt]
&       & \qqquad\qquad
          0 < l < q_k; v_i^j = w_k^l \} \\[\lngt]
& \cup  & \{ \tu{a_i^{2j+1,0},\nullConst} \mid 0 \leq i < m;
             0 \leq j < p_i \}  \\[\lngt]
& \cup  & \{ \tu{a_i^{2j+1,1},\rootConst} \mid 0 \leq i < m;
             0 \leq j < p_i \}   \\[\lngt]
\end{array}\]}
This completes the definition of $G$.  
\begin{claim}
The PCP instance has a solution iff there exists a
corresponder graph $G_0$ such that $G_0 \homs G$.
\end{claim}

$(\Longrightarrow):$ Assume that the PCP instance has a solution
$t_0,t_1,\ldots,t_{k-1}$.  Then
\[
   v_{t_0} v_{t_1} \ldots v_{t_{k-1}} = 
   w_{t_0} w_{t_1} \ldots w_{t_{k-1}}
\]
Let $u_0=l_0=0$,
\begin{eqnarray*}
n       &=& \sum_{j=0}^{k-1} |v_{t_j}| \\
u_{i+1} &=& \sum_{j=0}^{i} |v_{t_j}|, \qquad 0 \leq i < k-1 \\
l_{i+1} &=& \sum_{j=0}^{i} |w_{t_j}|, \qquad 0 \leq i < k-1 \\
\end{eqnarray*}
and let
\[
    G_0 = \CG(n,k,u_1,\ldots,u_{k-1},l_1,\ldots,l_{k-1})
\]
be a corresponder graph.  We construct a homomorphism $h$
from $G_0$ to $G$ as follows.  We map $C$-nodes of $G_0$
into $c$-nodes of $G$:
\[\begin{array}{rcll}
h(C_{2j}) &=& c_{2t_j}, & \qquad 0 \leq j < k \\
h(C_{2j+1}) &=& c_{2t_j+1}, & \qquad 0 \leq j < k \\
\end{array}\]
For $0 \leq f < n$, let $d_u(f)$ denote the largest index
$i$ such that $u_i \leq f$.  We map the $U$-nodes into
$a$ nodes as follows.  Consider a node $U_{2f}$.  Let
$i = d(f)$.  Then $U_{2f}$ is the even node that
represents the letter $v_i^{f-u_i}$ of the word $v_i$:
\[
    h(U_{2f}) = a_i^{2(f-u_i),0}, \qquad i=d_u(f), \
                                  0 \leq f < n
\]
The mapping of $U_{2f+1}$ is similar.  In this case
we also encode the information whether $L_{2f+1}$
has an $s_2$-edge from a $C$-node.
\[
    U_{2f+1} = a_i^{2(f-u_i)+1,\alpha_l(f)}, \qquad i=d_u(f), \
                                             0 \leq f < n
\]
where
\[
   \alpha_l(f) = \left\{\begin{array}{rl}
               1, & f \in \{ l_0,l_1,\ldots,l_{k-1} \} \\
               0, & \mbox{otherwise }
               \end{array}\right.
\]
The mapping of $L$-nodes is analogous.  Let $d_l(f)$ denote
the largest index $i$ such that $l_i \leq f$.  Then
\[
    L_{2f+1} = b_i^{2(f-u_i)+1,0}, \qquad i=d_u(f), \
                                   0 \leq f < n
\]
\[
    U_{2f} = b_i^{2(f-u_i),\alpha_u(f)}, \qquad i=d_u(f), \
                                         0 \leq f < n
\]
where
\[
   \alpha_u(f) = \left\{\begin{array}{rl}
               1, & f \in \{ u_0,u_1,\ldots,u_{k-1} \} \\
               0, & \mbox{otherwise }
               \end{array}\right.
\]
It is straightforward to verify that $h$ is indeed a
homomorphism.

$(\Longleftarrow):$ Assume that $G_0 \homs G$ where 
\[
    G_0 = \CG(n,k,u_1,\ldots,u_{k-1},l_1,\ldots,l_{k-1})
\]
is a corresponder graph and $h$ is a homomorphism from $G_0$
to $G$.  Because in graph $G$ all paths given by the regular
expression $1^{*}$ lead to $c_i$-nodes or $\nullConst$, we
conclude that each $C_j$ node is mapped to some $c_i$ node.
For $0 \leq j < k$ we define
\[
    t_j = i  \ \ \ifandonlyif \ \ h(C_{2j})=c_{2i}
\]
From the properties of homomorphism we derive
\[
    t_j = i \ \ \ifandonlyif \ \ h(C_{2j+1})=c_{2i+1}
\]
We will show that $t_j$ is a solution of the PCP instance.
Let $u_0=l_0=0$ and $u_{n+1}=l_{n+1}=n$.
Let
\[\begin{array}{l}
\tau(a_i^{2j,0}) = \tau(a_i^{2j+1,\alpha}) = v_i^j \\[\glngt]
\tau(b_i^{2j,\alpha}) = \tau(b_i^{2j+1,0}) = w_i^j \\
\end{array}\]
and $h' = \tau \circ h$.
By construction of $s_2$ in $G$ we have
\[
    h'(U_{2j}) = h'(U_{2j+1}) = h'(L_{2j}) = h'(L_{2j+1})
\]
for $0 \leq j < n$.  To prove
\[
   v_{t_0} v_{t_1} \ldots v_{t_{k-1}} = 
   w_{t_0} w_{t_1} \ldots w_{t_{k-1}}
\]
it therefore suffices to show
\begin{eqnarray}
\label{eqn:vInCorresponder}
v_{t_j} &=& h'(U_{2u_j}) h'(U_{2(u_j+1)}) \ldots 
    h'(U_{2(u_{j+1}-1)}) \quad \\
\label{eqn:wInCorresponder}
w_{t_j} &=& h'(L_{2l_j+1}) h'(L_{2(l_j+1)+1}) \ldots 
    h'(L_{2(l_{j+1}-1)+1}) \quad
\end{eqnarray}
for $0 \leq j < k$.  Let $t_j = i$.  Then
$h(C_{2j})=c_{2i}$.  We have $\tu{C_{2j},U_{2u_j}} \in s_2$
in the corresponder graph $G_0$.  On the other hand,
$\tu{c_{2i},a_i^{0,0}} \in s_2$ is the only $s_2$-outgoing
edge of $c_{2i}$ in $G$.  Therefore,
$h(U_{2u_j})=a_i^{0,0}$.  From this, we first conclude so
$h'(U_{2u_j})=v_i^0$.
Next, by construction of $s_1$ edges of $G$ and $G_0$ we
get
\[\begin{array}{ccl}
   h(U_{2(u_j+1)}) &=& a_i^{2,0} \\[\glngt]
   \ldots \\[\glngt]
   h(U_{2(u_j+p_i-1)}) &=& a_i^{2(p_i-1),0} \\
\end{array}\]
To establish~(\ref{eqn:vInCorresponder}) it suffices to show
$u_j+p_i = u_{j+1}$.  To see that the equality holds,
suppose first $u_j+p_i > u_{j+1}$.  Then
$h(U_{2u_{j+1}})=a_i^{2f,0}$ where $f > 0$.  Because $h$ is
a homomorphism, following $s_2$ edge once we conclude
$h(L_{2u_{j+1}})=b_g^{2r,0}$ for some $g$ and following
$s_2$ for the second time we obtain the contradiction
because in corresponder graph $s_2(L_{2u_{j+1}})=\rootConst$
but in $G$ we have $s_2(b_g^{2r,0})=\nullConst$.  Similarly,
suppose now $u_j+p_i < u_{j+1}$.  Then
$s_1(U_{2(u_j+p_i-1)}) \neq
\nullConst$, so let
$U_{2(u_j+p_i)}=s_1(s_1(U_{2(u_j+p_i-1)}))$.  Because
$u_j+p_i \notin \{ u_0,\ldots,u_{k-1} \}$, we get
\[
    s_2(s_2(U_{2(u_j+p_i)})) = \nullConst
\]
On the other hand, $h(U_{2(u_j+p_i)})=a_f^{0,0}$ for
some $f$, so
\[
    s_2(s_2(h(U_{2(u_j+p_i)}))) = \rootConst
\]
which is again a contradiction.  Therefore $u_j+p_i =
u_{j+1}$.  Showing~(\ref{eqn:wInCorresponder}) is analogous.
We conclude that $t_0,\ldots,t_{k-1}$ is a solution for
PCP instance.

Our claim is therefore true and satisfiability over
corresponder graphs is undecidable.
\end{proof}

\subsection{Defining Corresponder Graphs}

In this section construct graphs $P$ and $Q$
such that
\begin{equation} \label{eqn:negSat}
    G_0 \homs (P \land \lnot Q)
\end{equation}
iff $G_0$ is a corresponder graph.

When presenting the graphs $P,Q_0,\ldots,Q_{\negno}$ we use
the following conventions.  We use the label $r$ to denote
the root of the graph.  We label the edges of the relation
$s_1$ relation by $1$ and the edges of $s_2$ by $2$.  Note
that if a node has no outgoing edges, it would be useless in
the graph in terms of specifying a set of models $G_0$.
Every graph node will therefore have at least one outgoing
edge for every label.  However, in order to make the graph
sketches clearer, if a node $x$ has an outgoing edge with
label $a$ to every node in the graph, we will simply omit
all $a$ edges of node $x$ from the sketch.  In particular,
if a node has no outgoing edges in the graph sketch, it
means that its outgoing edges are unconstrained.  A
double-headed arrow from node $x$ to node $y$ with label $a$
denotes two single arrows, one from $x$ to $y$ and one from
$y$ to $x$, both labeled with $a$.  We do not show the edge
$\tu{\rootConst,\nullConst} \in s_2$ that is always present
in an orable graph.  We also do not show the edges
originating from $\nullConst$.  We will be free to display
$\nullConst$ several times in the same picture, all these
occurrences denote to the unique $\rootConst$ node in the
graph.

\begin{figure*}
\begin{center}
\mypic{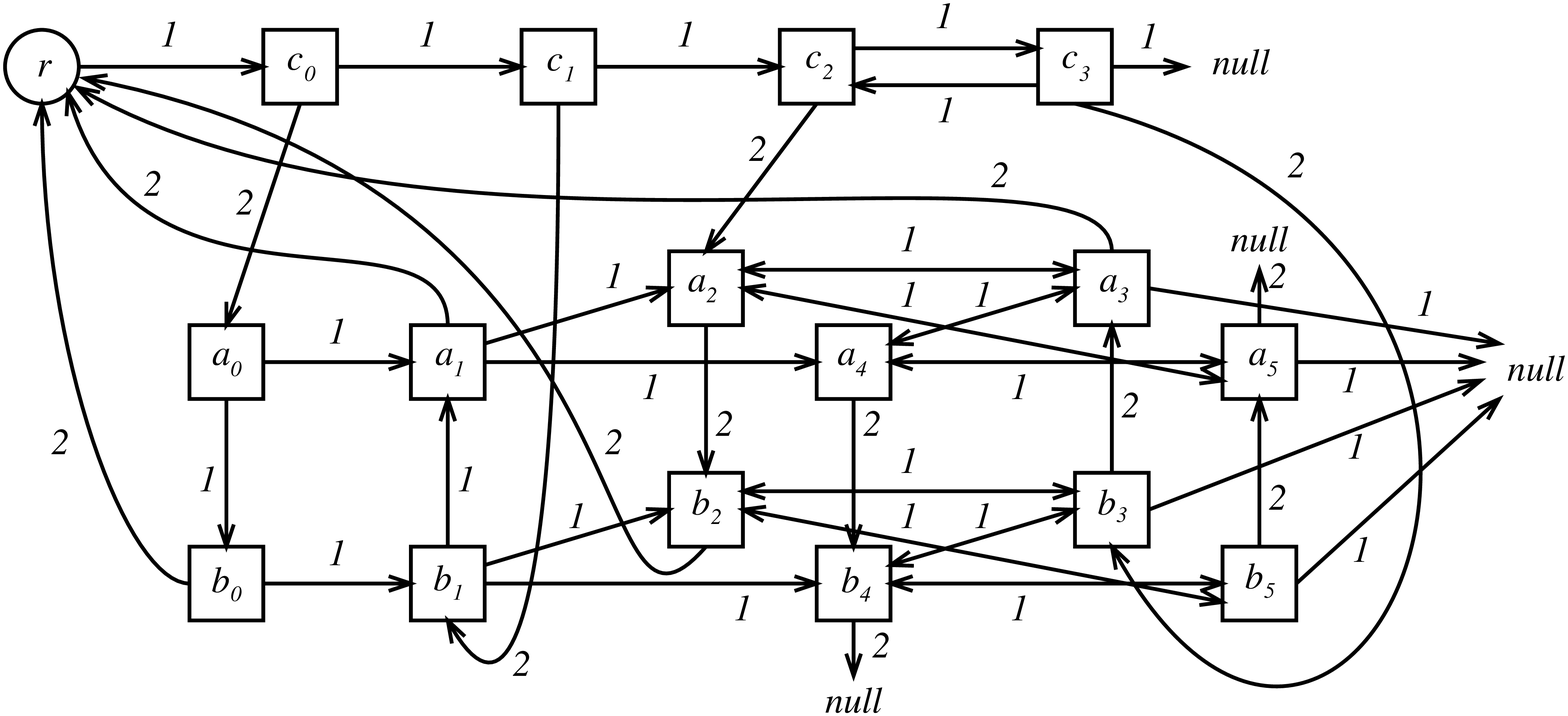}
\end{center}
\caption{Graph $P$\label{fig:graphP}}
\end{figure*}

\begin{figure*}
\begin{center}
\mypic{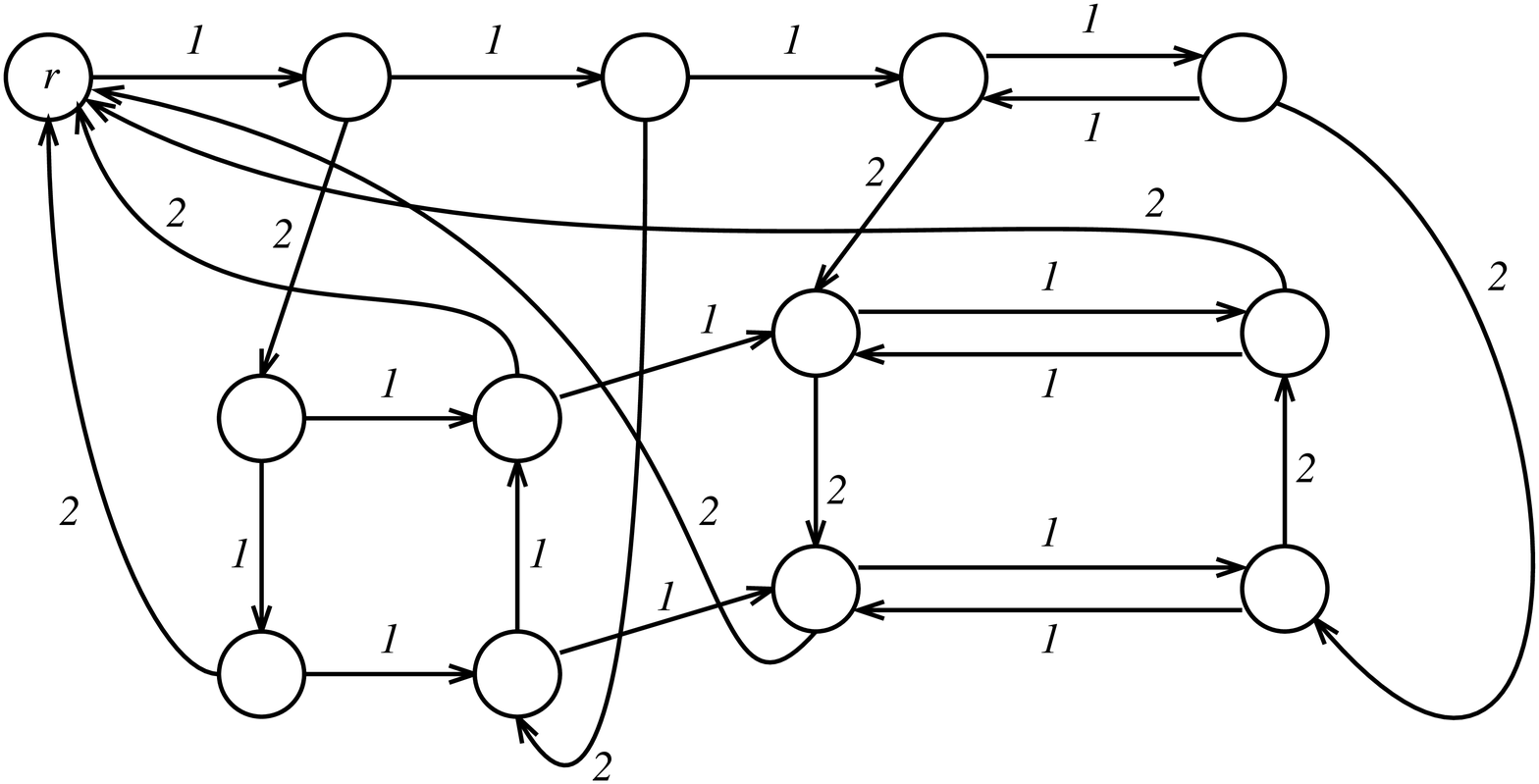}
\end{center}
\caption{A model of $P$ that is not a corresponder graph
\label{fig:graphCyclic}}
\end{figure*}

The graph $P$ in Figure~\ref{fig:graphP} is our first
approximation of a corresponder graph.  Unfortunately, $P$
allows some models that are not corresponder graphs, such as
the example in Figure~\ref{fig:graphCyclic}.  This is why we
introduce the graph $Q$.  The graph $Q$ appears in
$(\ref{eqn:negSat})$ negated and we design it to contain
models of $P$ that are not corresponder graphs.

\begin{figure}[thbp]
\begin{center}
\mypic{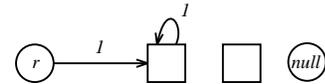}
\end{center}
\caption{Graph $Q_0$
\label{fig:graphQzero}}
\end{figure}

\begin{figure}[thbp]
\begin{center}
\mypic{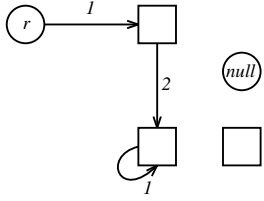}
\end{center}
\caption{Graph $Q_1$
\label{fig:graphQone}}
\end{figure}

\begin{figure}[thbp]
\begin{center}
\mypic{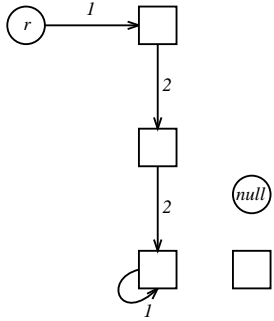}
\end{center}
\caption{Graph $Q_2$
\label{fig:graphQtwo}}
\end{figure}

\begin{figure}[thbp]
\begin{center}
\mypic{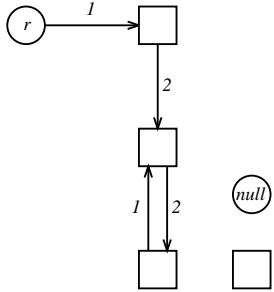}
\end{center}
\caption{Graph $Q_3$
\label{fig:graphQthree}}
\end{figure}

\begin{figure}[thbp]
\begin{center}
\mypic{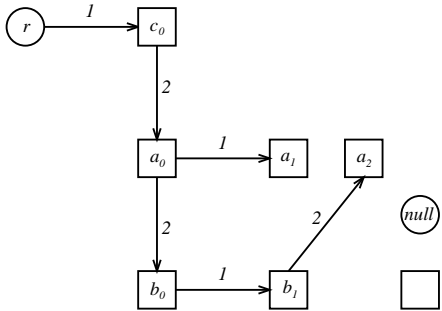}
\end{center}
\caption{Graph $Q_4$
\label{fig:graphQfour}}
\end{figure}

\begin{figure}[thbp]
\begin{center}
\mypic{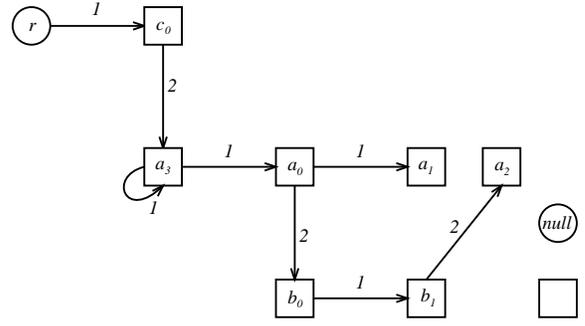}
\end{center}
\caption{Graph $Q_5$
\label{fig:graphQfive}}
\end{figure}

\begin{figure}[thbp]
\begin{center}
\mypic{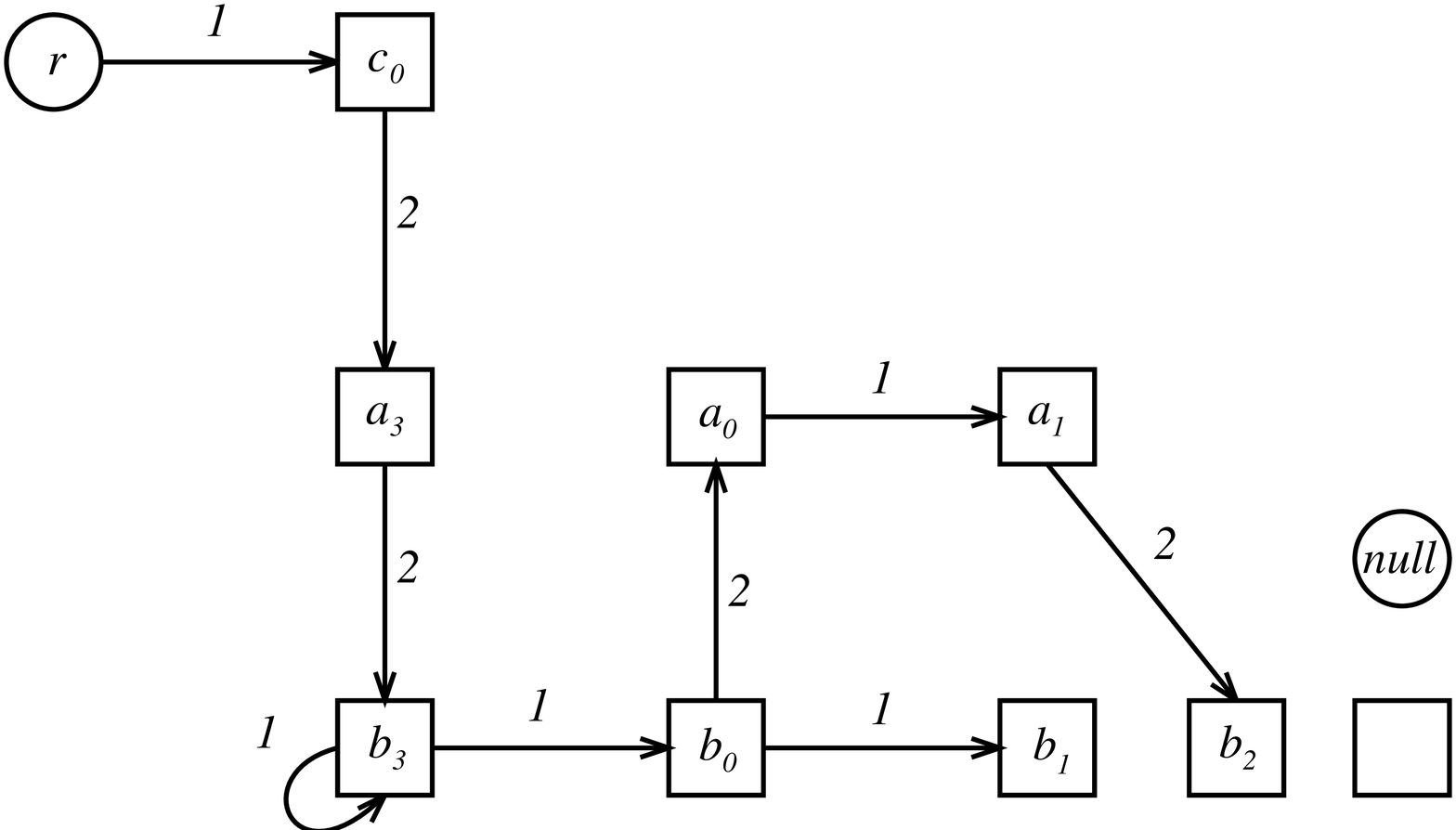}
\end{center}
\caption{Graph $Q_6$
\label{fig:graphQsix}}
\end{figure}

\begin{figure}[thbp]
\begin{center}
\mypic{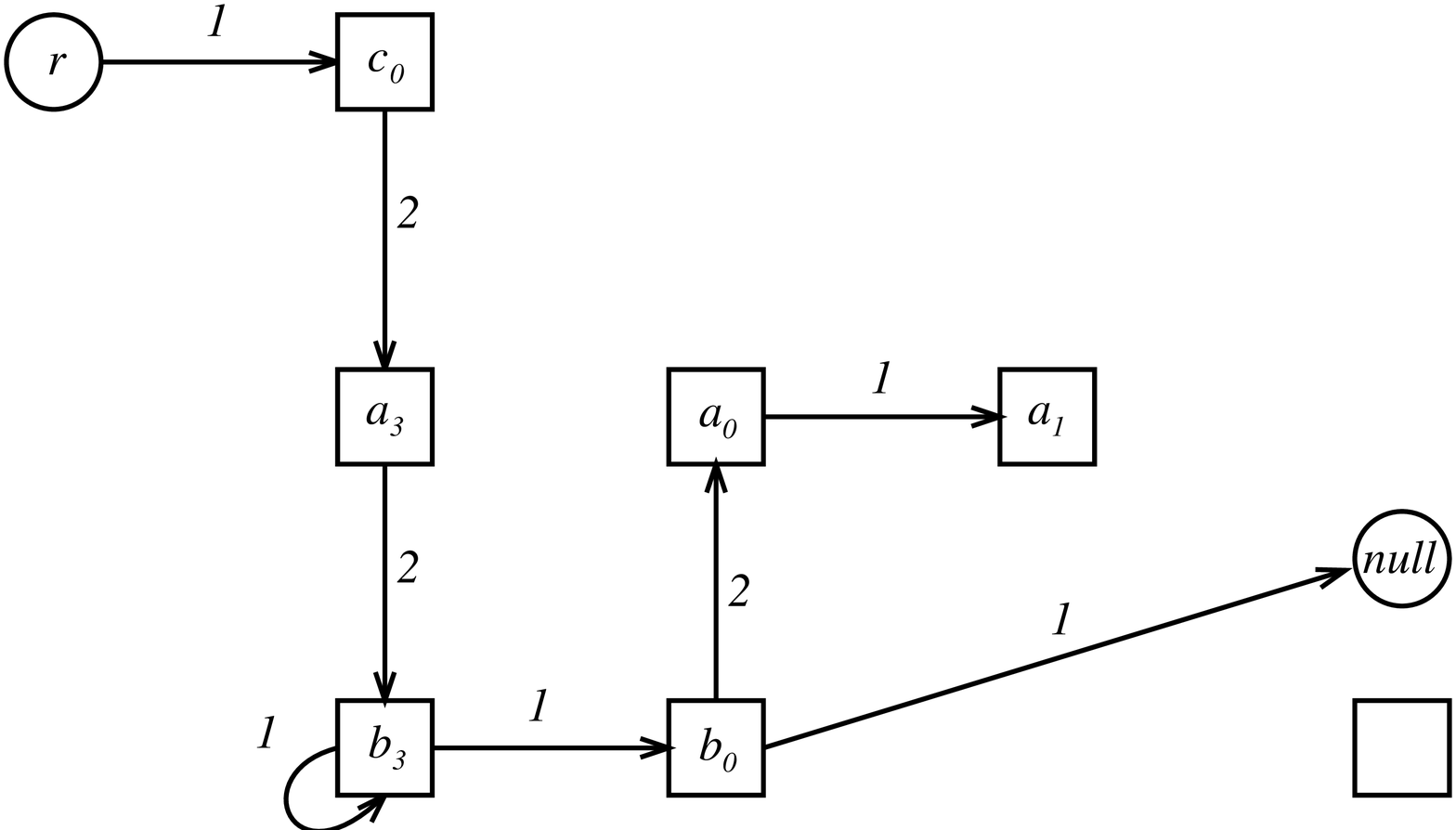}
\end{center}
\caption{Graph $Q_7$
\label{fig:graphQseven}}
\end{figure}

\begin{figure}[thbp]
\begin{center}
\mypic{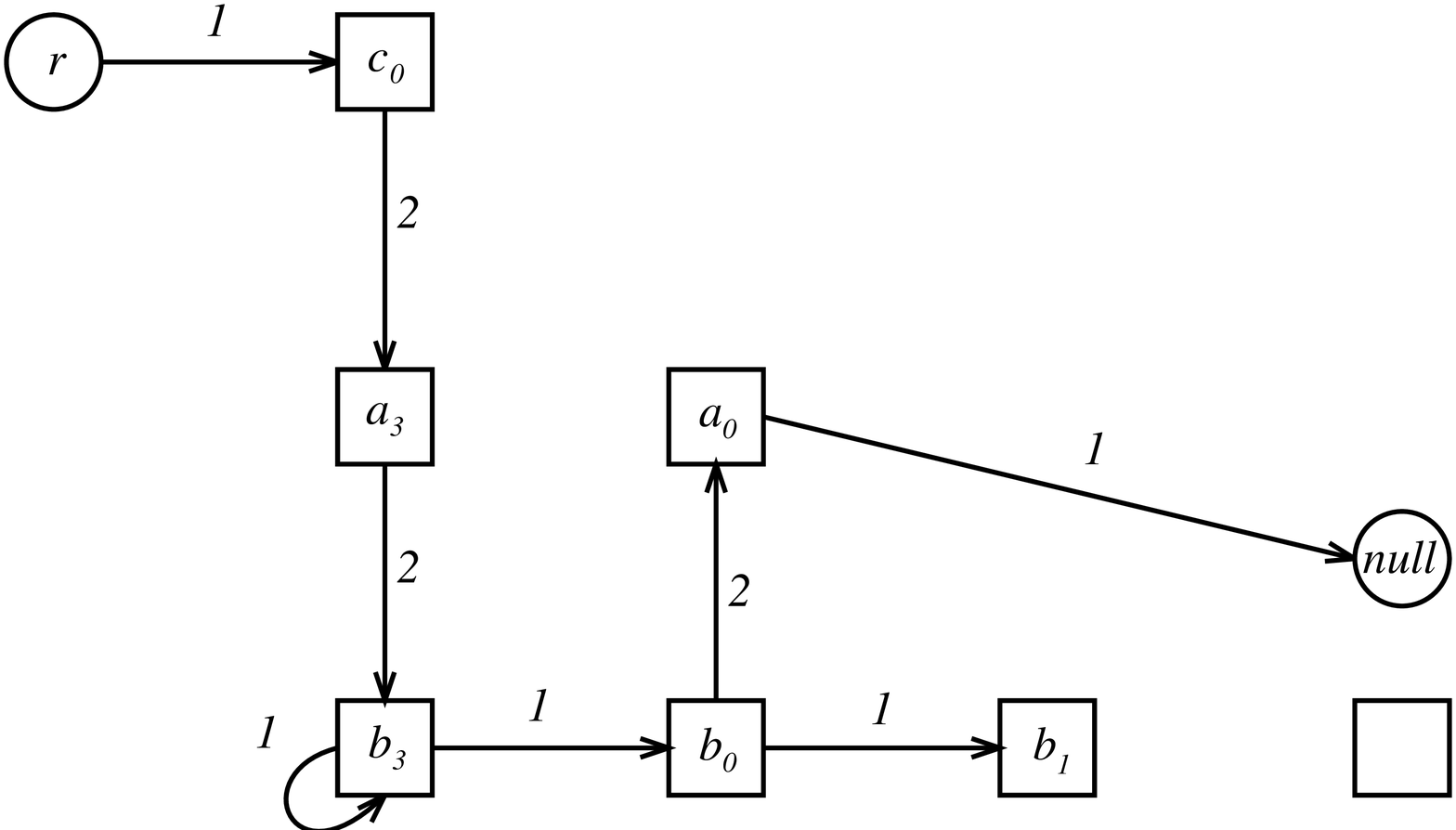}
\end{center}
\caption{Graph $Q_8$
\label{fig:graphQeight}}
\end{figure}

\begin{figure}[thbp]
\begin{center}
\mypic{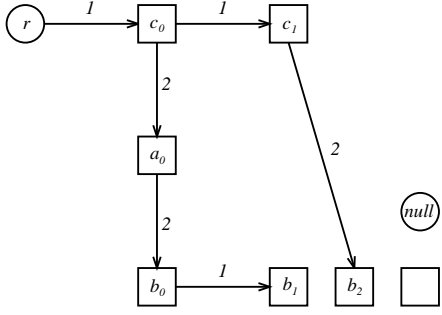}
\end{center}
\caption{Graph $Q_9$
\label{fig:graphQnine}}
\end{figure}

\begin{figure}[thbp]
\begin{center}
\mypic{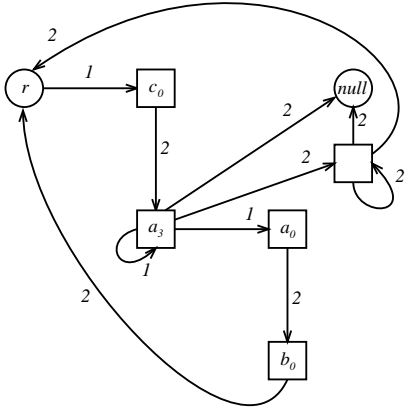}
\end{center}
\caption{Graph $Q_{10}$
\label{fig:graphQten}}
\end{figure}

\begin{figure}[thbp]
\begin{center}
\mypic{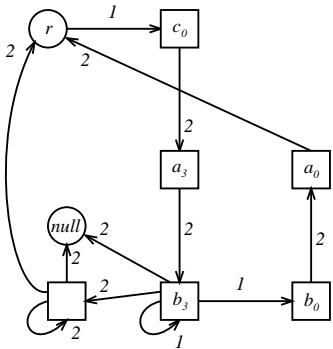}
\end{center}
\caption{Graph $Q_{11}$
\label{fig:graphQeleven}}
\end{figure}

\begin{figure*}[thbp]
\begin{center}
\mypic{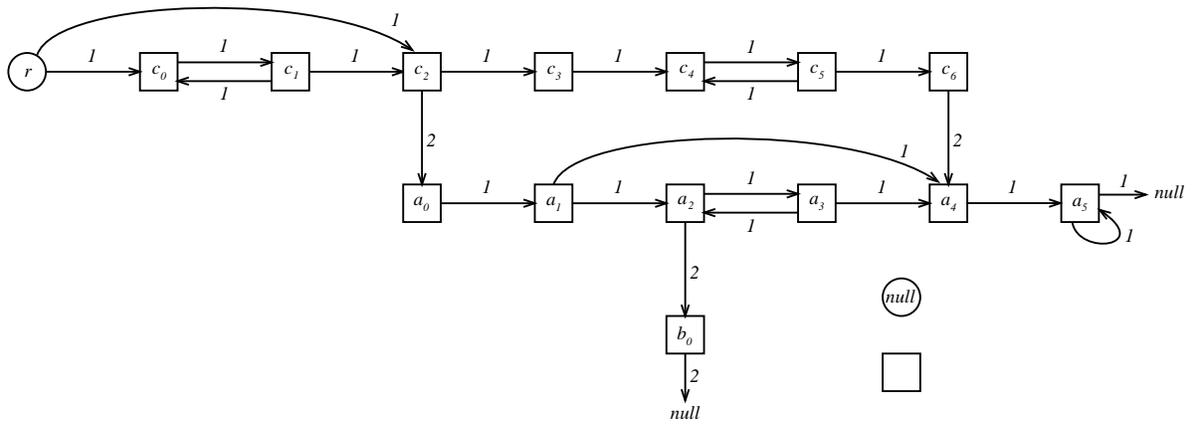}
\end{center}
\caption{Graph $Q_{12}$
\label{fig:graphQtwelve}}
\end{figure*}

\begin{figure*}[thb]
\begin{center}
\mypic{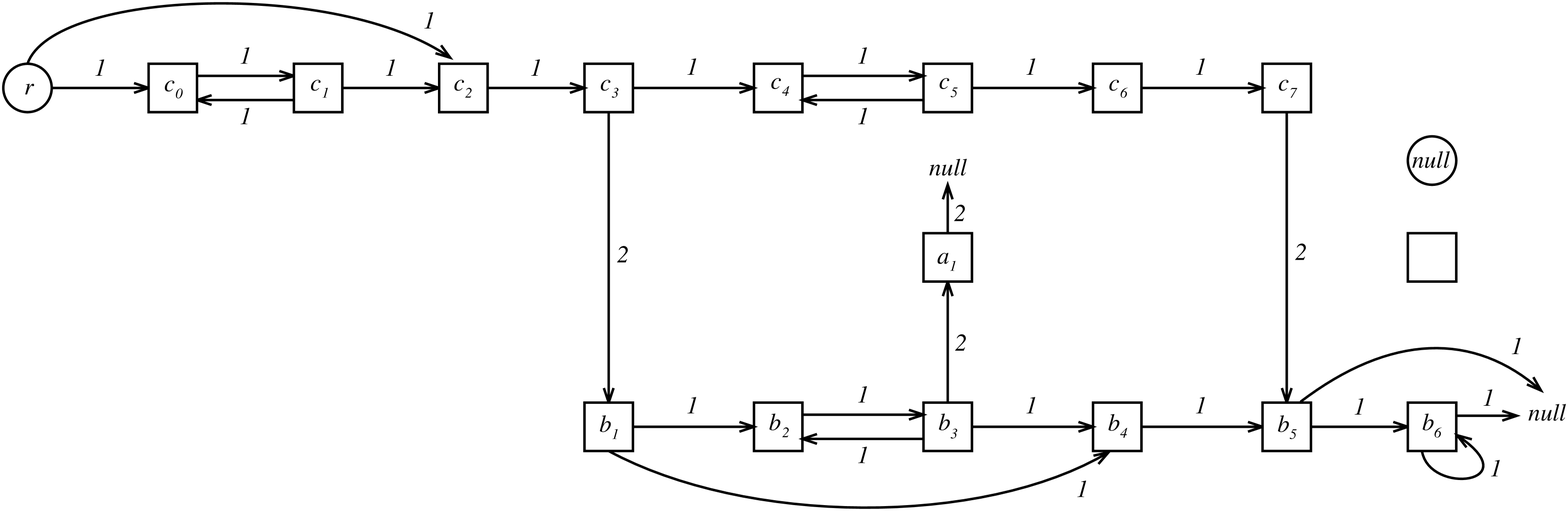}
\end{center}
\caption{Graph $Q_{13}$
\label{fig:graphQthirteen}}
\end{figure*}

We construct $Q$ as a sum of orable graphs:
\[
    Q = Q_0 + Q_1 + \cdots + Q_{\negno}
\]
The idea behind the construction of these graphs comes from
the proof of Proposition~\ref{prop:corrEncoding}; we now
give only an informal overview of the graphs.  The graphs
$Q_0$ (Figure~\ref{fig:graphQzero}) $Q_1$
(Figure~\ref{fig:graphQone}), $Q_2$
(Figure~\ref{fig:graphQtwo}), and $Q_3$
(Figure~\ref{fig:graphQthree}) eliminate certain cycles from
the set of models of $P$.  The graphs $Q_4$
(Figure~\ref{fig:graphQfour}), $Q_5$
(Figure~\ref{fig:graphQfive}), $Q_6$
(Figure~\ref{fig:graphQsix}), and $Q_9$
(Figure~\ref{fig:graphQnine}) ensure that different paths in
the graph lead to the same object.  The graphs $Q_7$
(Figure~\ref{fig:graphQseven}) and $Q_8$
(Figure~\ref{fig:graphQeight}) ensure that there is the same
number of $U$ and $L$-nodes in a model of $P$.  The graphs
$Q_{10}$ (Figure~\ref{fig:graphQten}) and $Q_{11}$
(Figure~\ref{fig:graphQeleven}) ensure that $U$ or $L$ nodes
have an $s_2$ edge to $\rootConst$ iff the $U$ or $L$ node
in the same column has an $s_2$-edge from a $C$-node.  The
graphs $Q_{12}$ (Figure~\ref{fig:graphQtwelve}) and $Q_{13}$
(Figure~\ref{fig:graphQthirteen}) ensure that a $C$-node
that is later in the $C$-list has an edge to a node that is
later in the $U$ or $L$ list.  Finally, graphs $Q_{14}$
(Figure~\ref{fig:graphQfourteen}), $Q_{15}$
(Figure~\ref{fig:graphQfifteen}) and $Q_{16}$
(Figure~\ref{fig:graphQsixteen}) ensure that $C$-nodes have
$s_2$ edges only to $U$ and $L$-nodes, and that an $L$ or
$U$ node can only have an edge to $\rootConst,\nullConst$, a
$U$-node, or an $L$-node.

\noindent We can now show the key step in the undecidability
proof for the implication of graph constraints.
\begin{proposition} \label{prop:corrEncoding}
\begin{equation}
    G_0 \homs (P \land \lnot Q)
\end{equation}
iff $G_0$ is a corresponder graph.
\end{proposition}
\begin{proof}
$(\Longleftarrow):$ Let $G_0$ be a corresponder graph
\[
    G_0 = \CG(n,k,u_1,\ldots,u_{k-1},l_1,\ldots,l_{k-1})
\]
We show that $G_0 \homs P$ and for all $0 \leq i \leq
\negno$, it is not the case that $G_0 \homs Q_i$.

$(G_0 \homs P):$  Define homomorphism $h$ from $G_0$ to $P$
as follows.
\[\begin{array}{rcl}
  h(C_0) &=& c_0 \mnls
  h(C_1) &=& c_1 \mnls
  h(C_{2j+2}) &=& c_2, \qquad 0 \leq j < k-1 \mnls
  h(C_{2j+3}) &=& c_3, \qquad 0 \leq j < k-1 \mnls
  h(U_0) &=& a_0 \mnls
  h(U_1) &=& a_1 \mnls
  h(U_{2j+2}) &=& \left\{\begin{array}{rl}
                  a_2, & j+1 \in \{ u_1,\ldots,u_{k-1} \} \\
                  a_4, & \mbox{otherwise} \\
                         \end{array}\right. \\[3ex]
  h(U_{2j+3}) &=& \left\{\begin{array}{rl}
                  a_3, & j+1 \in \{ l_1,\ldots,l_{k-1} \} \\
                  a_5, & \mbox{otherwise} \\
                         \end{array}\right. \mnl
  h(L_0) &=& b_0 \mnl
  h(L_1) &=& b_1 \mnl
  h(L_{2j+2}) &=& \left\{\begin{array}{rl}
                  b_2, & j+1 \in \{ u_1,\ldots,u_{k-1} \} \\
                  b_4, & \mbox{otherwise} \\
                         \end{array}\right. \\[3ex]
  h(L_{2j+3}) &=& \left\{\begin{array}{rl}
                  b_3, & j+1 \in \{ l_1,\ldots,l_{k-1} \} \\
                  b_5, & \mbox{otherwise} \\
                         \end{array}\right. \mnl
\end{array}\]
It is straightforward to verify that $h$ is indeed a
homomorphism, so $G_0 \homs P$.

$(\lnot\ G_0 \homs Q_0):$ Apply
Proposition~\ref{prop:regexpTest} with $e = 1^{*}$.

$(\lnot\ G_0 \homs Q_1):$ Apply
Proposition~\ref{prop:regexpTest} with $e = 121^{*}$.

$(\lnot\ G_0 \homs Q_2):$ Apply
Proposition~\ref{prop:regexpTest} with $e = 1221^{*}$.

$(\lnot\ G_0 \homs Q_3):$ Apply
Proposition~\ref{prop:regexpTest} with $e = 12(21)^{*}$.

$(\lnot\ G_0 \homs Q_4):$ Suppose $h$ is a homomorphism from
$G_0$ to $Q_4$.  By mapping the path
\[
   \rootConst,1,C_0,2,U_0,1,U_1
\]
we conclude $h(U_1)=a_1$.  By
mapping the path 
\[
   \rootConst,1,C_0,2,U_0,2,L_0,1,L_1,2,U_1
\]
we conclude $h(U_1)=a_2$, which is a contradiction.

$(\lnot\ G_0 \homs Q_5):$ Suppose $h$ is a homomorphism from
$G_0$ to $Q_5$.  By mapping the slice in $121^{*}$ with $h$
we conclude that there exists a node $U_j$ in $G_0$ such that
$h(U_j)=a_0$.  Now as in the previous case we get that
$h(U_{j+1})=a_1$ and $h(U_{j+1})=a_2$, which is a
contradiction.

$(\lnot\ G_0 \homs Q_6):$ Similarly to the previous case,
map the slice in $1221^{*}$ to conclude that for some node
$L_j$ we have $h(L_j)=b_0$ and then obtain a contradiction.

$(\lnot\ G_0 \homs Q_7):$ Similarly to the previous cases,
suppose $h$ is a homomorphism from $G_0$ to $Q_7$.  By
mapping the slice in $1221^{*}$ via $h$, we conclude that there
exists a node $L_j$ in $G_0$ such that $h(L_j)=b_0$.  By
construction of $Q_7$ it must be $s_1(L_j)=\nullConst$.
Furthermore, $h(U_j)=a_0$ and $s_1(U_j)\neq \nullConst$.
This is a contradiction with the fact that $G_0$ is a
corresponder graph.

$(\lnot\ G_0 \homs Q_8):$ This fact is analogous to the
previous one.

$(\lnot\ G_0 \homs Q_9):$ Suppose $h$ is a homomorphism from
$G_0$ to $Q_9$.  As in the case $\lnot\ G_0 \homs Q_4$ we
conclude $h(L_1)=b_1$ and $h(L_1)=b_2$, a contradiction.

$(\lnot\ G_0 \homs Q_{10}):$ Suppose $h$ is a homomorphism
from $G_0$ to $Q_{10}$.  As in case $\lnot\ G_0 \homs Q_5$,
for some node $U_j$ in $G_0$ we have $h(U_j)=a_0$.  Next it
follows $h(L_j)=b_0$ and $s_2(L_j)=\rootConst$, which
implies $j=2u_r$ where $0 \leq r < k$.  Therefore
$s_2(C_{2r})=U_j$.  But this is in contradiction with the
fact that $h(U_j)=a_0$ and $a_0$ has no incoming
$s_2$-edges in $Q_{10}$.

$(\lnot\ G_0 \homs Q_{11}):$ This case is analogous to the
previous one.

$(\lnot\ G_0 \homs Q_{12}):$ Suppose $h$ is a homomorphism
from $G_0$ to $Q_{10}$.  By mapping the slice in $1^{*}$
from $G_0$ to $G$ we conclude that there exists a node
$C_{2j}$ in $G_0$ such that $h(C_{2j})=c_2$ and a node
$C_{2i}$ where $i \geq j+2$ such that $h(C_{2i})=c_6$.
Since $s_2(c_2)=a_0$, we conclude $h(U_{2u_j})=a_0$.
By mapping the path
\[
   U_{2j},1,U_{2j+1},1,U_{2j+2},1,\ldots,1,\nullConst
\]
with homomorphism $h$, we conclude that there exists
be some $U$-node with even index that is mapped to $a_4$.
So let $U_{2(u_j+t)}$ be such node with the least index.  Then
\begin{equation} \label{eqn:firstNodeMaps}
    h(U_{2(u_j+t)})=a_4
\end{equation}
and
\[
   h(U_{2(u_j+r)}) = a_2
\]
for $1 \leq r < t$.   Let $1 \leq r < t$.  Then
$h(L_{2(u_j+r)}) = b_0$ so $s_2(L_{2(u_j+r)})=\nullConst$,
which means that $u_j+r < u_{j+1}$ for all $1 \leq r < t$,
so $u_j+t \leq u_{j+1} < u_i$.  Therefore
\[
   s_2(C_{2i}) \neq U_{2(u_j+t)}
\]
$h$ is a homomorphism and $h(C_{2i})=c_6$, so
\begin{equation} \label{eqn:secondNodeMaps}
    h(U_{2u_i}) = a_4
\end{equation}
The corresponder graph $G_0$ contains a path in $1^{*}$ from
$U_{2(u_j+t)}$ to $U_{2u_i}$ where $U_{2(u_j+t)}$ and
$U_{2u_i}$ are distinct nodes because $u_j+t < u_i$.
Because $h(U_{2(u_j+t)})=h(U_{2u_i})=a_4$, there exists a
cyclic $s_1$-path from $a_4$ to $a_4$ in $Q_{12}$, a
contradiction with the definition of $Q_{12}$.

$(\lnot\ G_0 \homs Q_{13}):$ This case is analogous to the
previous one.

$(\lnot\ G_0 \homs Q_{14}):$ Suppose $h$ is a homomorphism
from $G_0$ to $Q_{14}$.  Then $h(C_i)=c_2$ for some $i$.
Let $S = s_2(C_i)$.  Then $S=U_j$ or $S=L_j$ for some $j$
and $h(S) = a_2$.  On the other hand, by mapping the slices
$121^{*}$ and $1221^{*}$, we conclude that all $U$-nodes are
mapped to $a_0$ and $a_1$ whereas all $L$-nodes are mapped
to $b_0$ and $b_1$.  This is a contradiction with
$h(S)=a_2$.

$(\lnot\ G_0 \homs Q_{15}):$ Suppose $h$ is a homomorphism
from $G_0$ to $Q_{15}$.  Then for some node $L_j$ we have
$h(L_j)=b_2$ and therefore $h(U_j)=a_2$.  On the other hand
by mapping the slice $121^{*}$ we conclude that all
$U$-nodes are mapped to $a_0$ and $a_1$, which is a
contradiction.

$(\lnot\ G_0 \homs Q_{16}):$ This case is analogous to the
previous one.

\begin{figure}[thb]
\begin{center}
\mypic{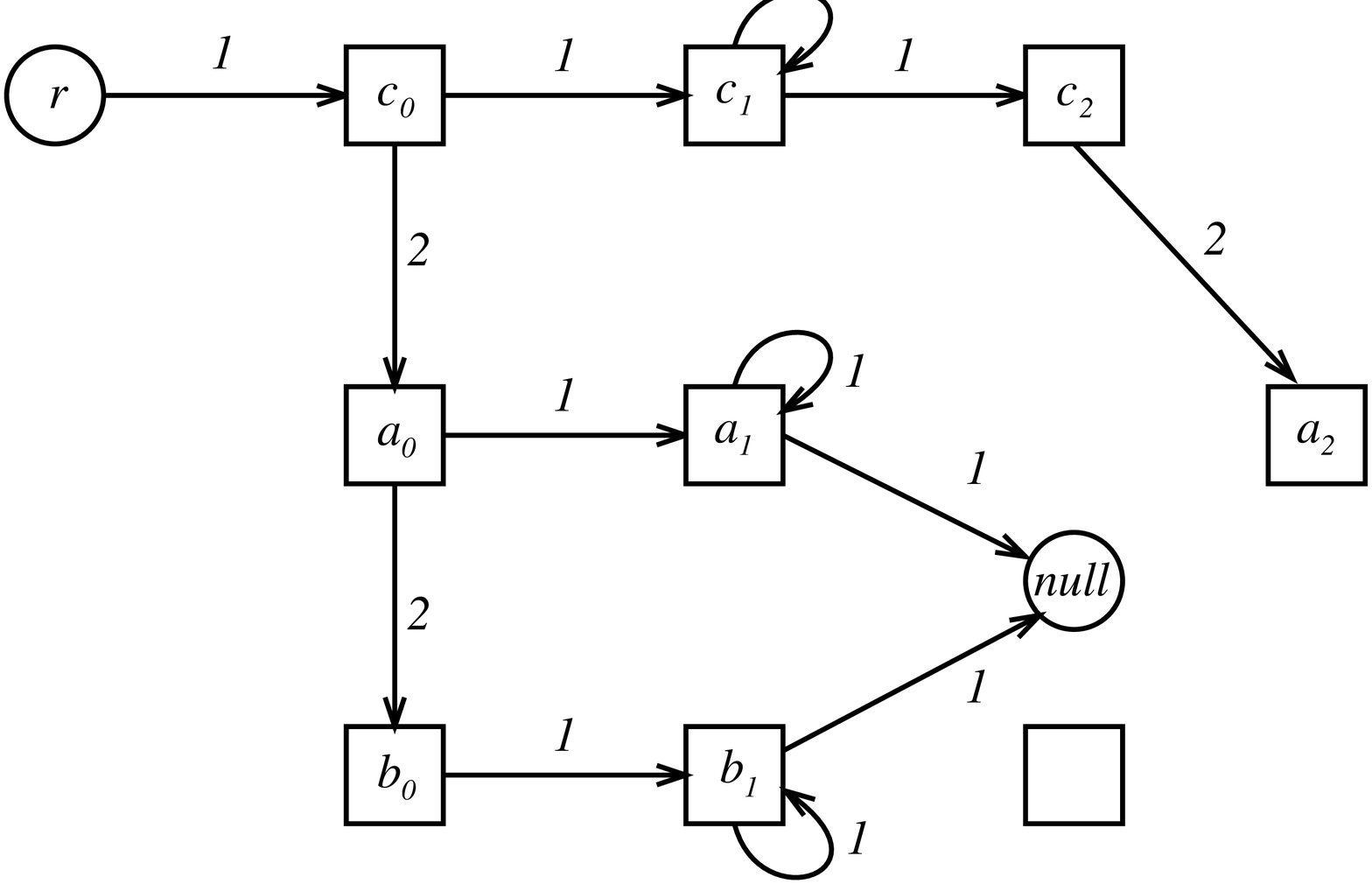}
\end{center}
\caption{Graph $Q_{14}$
\label{fig:graphQfourteen}}
\end{figure}

\begin{figure}[thb]
\begin{center}
\mypic{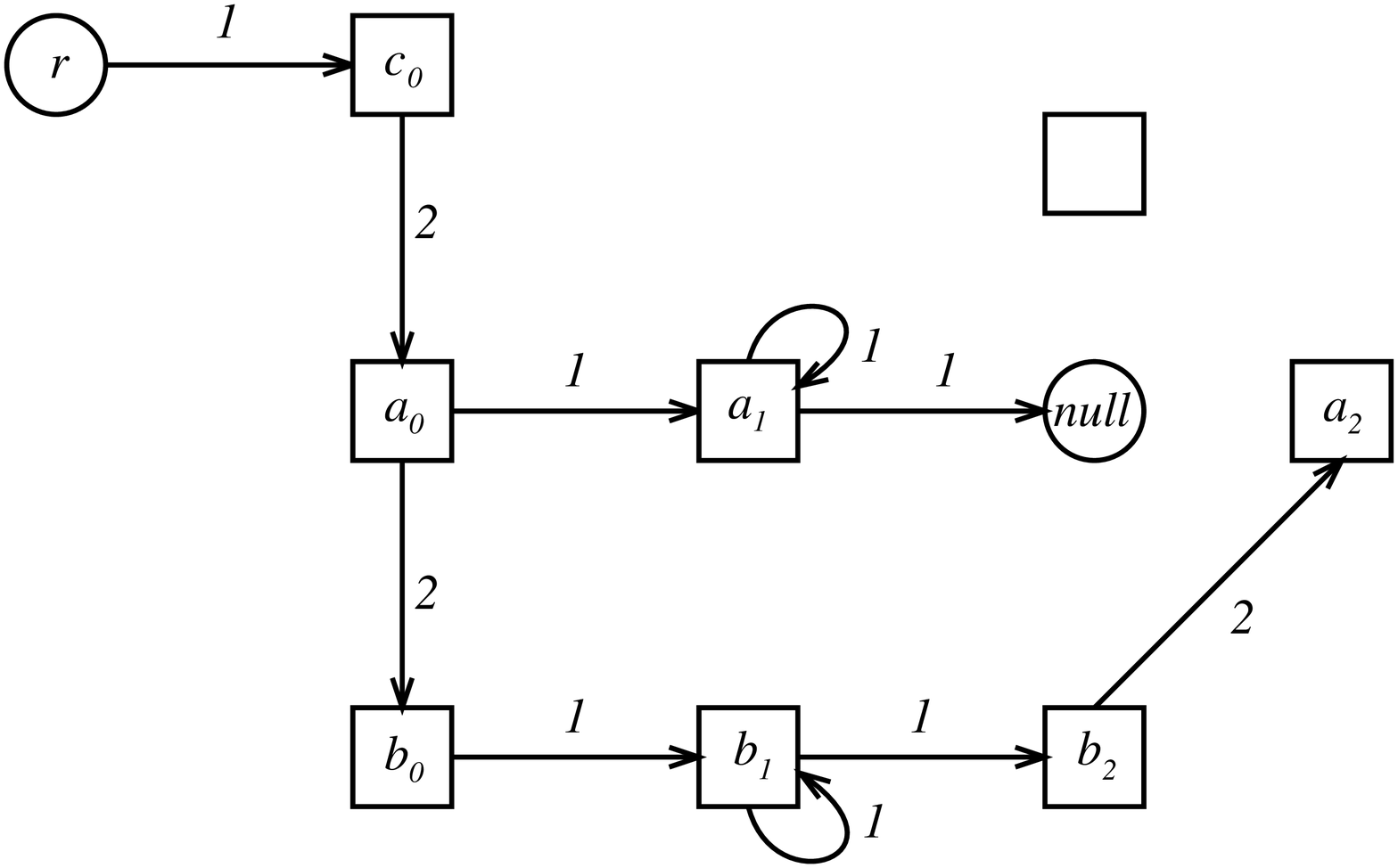}
\end{center}
\caption{Graph $Q_{15}$
\label{fig:graphQfifteen}}
\end{figure}

\begin{figure}[thb]
\begin{center}
\mypic{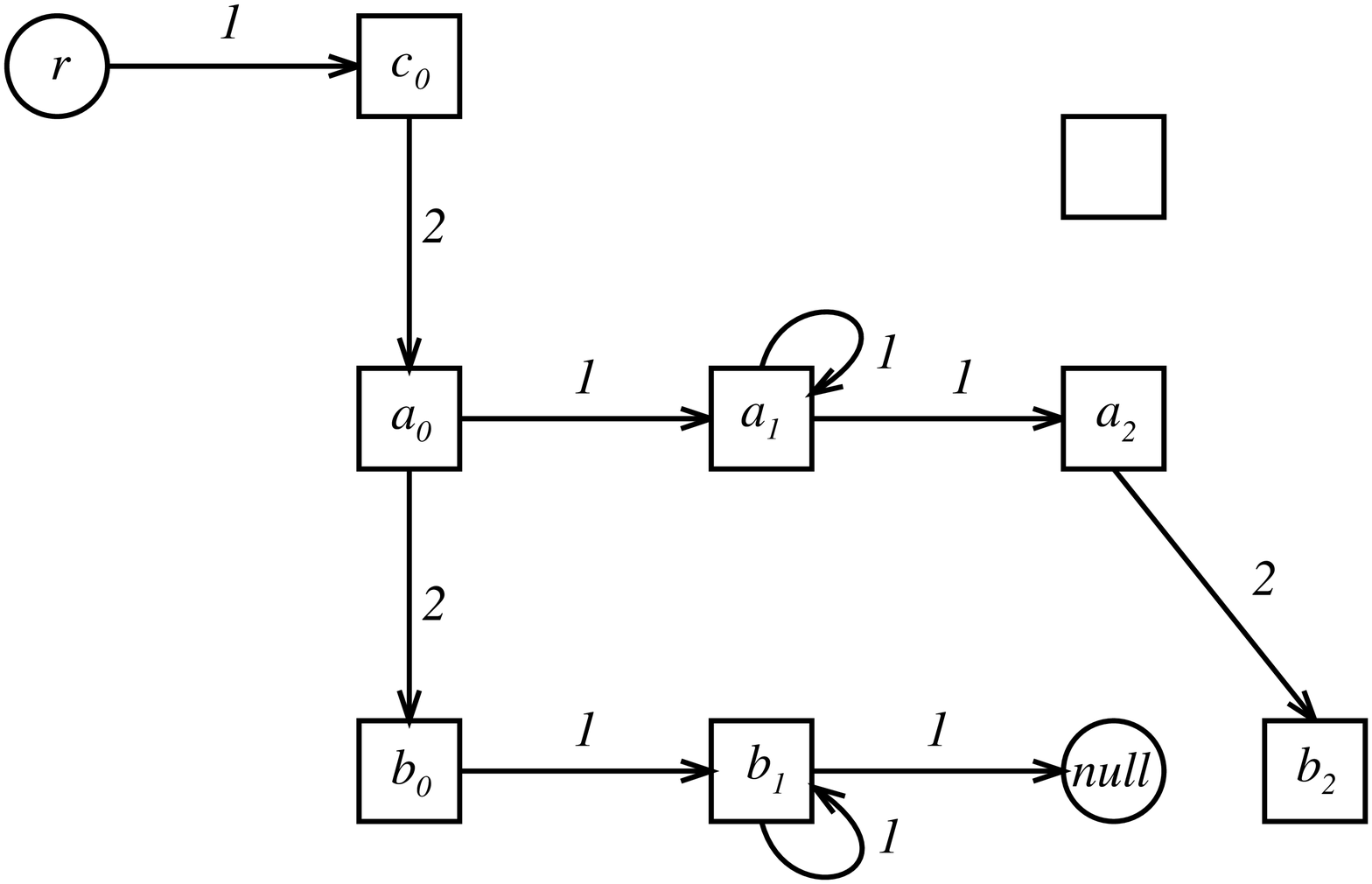}
\end{center}
\caption{Graph $Q_{16}$
\label{fig:graphQsixteen}}
\end{figure}

$(\Longrightarrow):$ Let
\[
    G_0 = \tu{V,s_1,s_2,\nullConst,\rootConst}
\]
Assume that $G_0 \homs P$ and for all $0 \leq i
\leq
\negno$ it is not the case that $G_0 \homs Q_i$.  We
will show that $G_0$ is a corresponder graph.  Let
\[\begin{array}{rcl}
  C_0 &=& s_1(\rootConst)  \mnl
  C_1 &=& s_1(C_0) \mnl
  C_2 &=& s_1(C_1) \mnl
  &\cdots&
\end{array}\]
Then for all $C_i \geq 0$ we have $C_i \neq \rootConst$,
because $G_0 \homs P$.  

We claim that there must exist $t$ such that
$C_t=\nullConst$ and $C_{t-1} \neq
\nullConst$.  Suppose the claim is false.  Because the
graph is finite, the nodes $C_i$ form a cycle with $s_1$
edges: there exist $i_1,i_2$ such that $0 < i_1 < i_2$,
$C_{i_1}=C_{i_2}$ and $C_{i_1} \neq \nullConst$.  We can
then show $G_0 \homs Q_0$, a contradiction.

Let $t$ be the smallest index such that $C_t=\nullConst$.
Then $t=2k$ for some $k \geq 2$ because $G_0 \homs P$.  The
nodes $\rootConst,\nullConst,C_0,\ldots,C_{2k-1}$ are all
distinct.

Next, consider the sequence
\[\begin{array}{rcl}
 U_0 &=& s_2(C_0) \mnl
 U_1 &=& s_1(U_0) \mnl
 U_2 &=& s_1(U_1) \mnl
 &\cdots&
\end{array}\]
Because $\lnot\ G_0 \homs Q_1$ and $G_0 \homs P$ there must
exist some $n_1$ such that $U_{2n_1}=\nullConst$ and
\[\begin{array}{r@{\,}l}
   U_i \notin \{\nullConst,\rootConst\} 
       & \cup \, \{ C_0,\ldots,C_{2k-1} \} \mnl
       & \cup \, \{ U_0,\ldots,U_{i-1} \}
\end{array}\]
for $0 \leq i < 2n_1$.  Because $\lnot G_0 \homs Q_2$ and
$G_0 \homs P$, considering the sequence
\[\begin{array}{rcl}
 L_0 &=& s_2(U_0) \mnl
 L_1 &=& s_1(L_0) \mnl
 L_2 &=& s_1(L_1) \mnl
 &\cdots&
\end{array}\]
we conclude there must exist $n_2$ such that
$L_{2n_2}=\nullConst$ and
\[\begin{array}{r@{\,}l}
   L_i \notin \{\nullConst,\rootConst\} 
       & \cup \, \{ C_0,\ldots,C_{2k-1} \} \mnl
       & \cup \, \{ U_0,\ldots,U_{2n_1-1} \} \mnl
       & \cup \, \{ L_0,\ldots,L_{i-1} \} \mnl
\end{array}\]
for $0 \leq i < 2n_2$.  Let
\[\begin{array}{r@{\,}l}
   V_0 = \{\nullConst,\rootConst\} 
       & \cup \, \{ C_0,\ldots,C_{2k-1} \} \mnl
       & \cup \, \{ U_0,\ldots,U_{2n_1-1} \} \mnl
       & \cup \, \{ L_0,\ldots,L_{2n_2-1} \} \mnl
\end{array}\]

\begin{figure*}
\begin{center}
\mypic{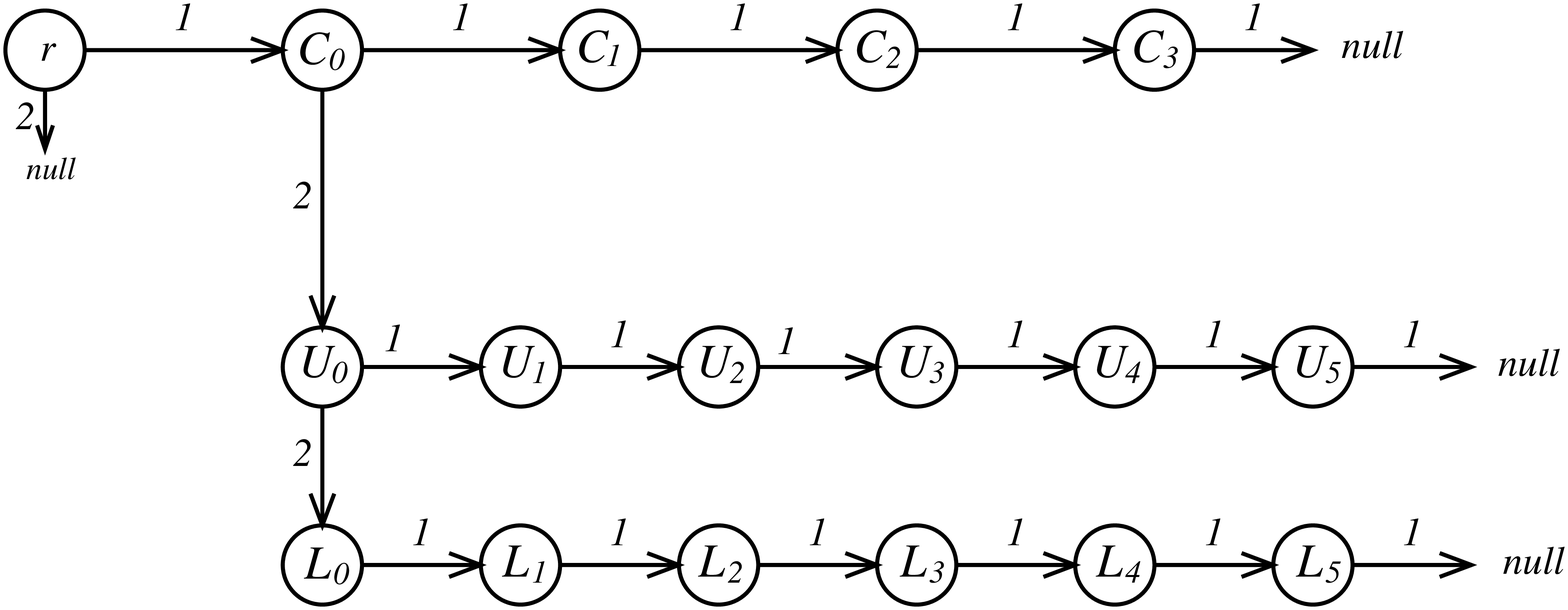}
\end{center}
\caption{Graph $G_0$ after identifying the nodes
\label{fig:corresponderToBe}}
\end{figure*}

\noindent
Figure~\ref{fig:corresponderToBe} shows the shape of the
portion of $G_0$ identified so far.  By construction,
\[
    s_1[V_0] \subseteq V_0
\]
In the sequel we will show that $s_2[V_0] \subseteq V_0$
holds as well.  By definition of heap, all nodes in $G_0$
are reachable from $\rootConst$, which will imply $V
\setminus V_0 =
\emptyset$.  We will also show that $n_1=n_2$ and that $G_0$
satisfies the invariants that make it isomorphic to a
corresponder graph.

We first observe that $s_2(C_1)=L_1$.  Indeed, suppose that
$s_2(C_1) \neq L_1$.  From $G_0 \homs P$ follows $s_2(C_1)
\notin \{\nullConst,\rootConst\}$.  We can then
show $G_0 \homs Q_9$, which is a contradiction.

We now show that the $s_2$ edges between $U$-nodes and
$L$-nodes form a $(2\times n)$-grid where $n=n_1=n_2$.  First
we observe $s_2(L_1)=U_1$, otherwise we would have $G_0
\homs Q_4$.  Next we claim that every non-null $s_2$ edge
originating from a $U$-node terminates at an $L$-node.
Suppose $s_2(U_j)$ is not an $L$-node.  It cannot be a
$C$-node or a $U$-node because $G_0 \homs P$.  The only
remaining possibility is that $s_2(U_j)$ is a node outside
$V_0$.  But then $G_0 \homs Q_{16}$, a contradiction.
Similarly, because $\lnot\ G_0 \homs Q_{15}$, every non-null
$s_2$ edge of an $L$-node terminates at a $U$-node.
Finally, we claim that for all $j \geq 0$, either
$U_{2j}=L_{2j}=\nullConst$, or all of the following holds:
\begin{itemize}
\item $\nullConst \notin \{ U_{2j}, L_{2j}, U_{2j+1}, L_{2j+1} \}$
\item $s_2(U_{2j})=L_{2j}$
\item $s_2(L_{2j+1})=U_{2j+1}$
\end{itemize}
We have already established the claim for $j=0$.  Suppose
the claim does not hold for all $j$.  Consider the least
$j>0$ for which the claim does not hold.  Then one of the nodes
$U_{2j}$, $L_{2j}$ is not $\nullConst$.  Assume $U_{2j} \neq
\nullConst$ and $L_{2j} = \nullConst$.  Then $G_0 \homs
Q_7$, a contradiction.  Similarly, if $U_{2j} = \nullConst$
and $L_{2j} \neq \nullConst$, then $G_0 \homs Q_8$, again a
contradiction.  So $U_{2j} \neq \nullConst$ and $L_{2j} \neq
\nullConst$.  Then from $G_0 \homs P$ follows $U_{2j+1} \neq
\nullConst$ and $L_{2j} \neq \nullConst$.
From the previous discussion and $G_0 \homs P$ we conclude
\[
    s_2(U_{2j}) = L_{2i}
\]
for some $i \geq 0$.  We want to show $i=j$.  Suppose $i <
j$.  Then there is a cycle $p$ starting at $U_0$ such that
$\word(p) \in (21)^{*}$.
But then $G_0 \homs Q_3$, a contradiction.
Now suppose $i > j$.  Then $G_0 \homs Q_6$, a contradiction.
Therefore $i=j$ and $s_2(U_{2j})=L_{2j}$.  We similarly
establish $s_2(L_{2j+1})=U_{2j+1}$ using the fact $\lnot\
G_0 \homs Q_5$.  This establishes our claim for all $j \geq
0$.  We conclude that $n_1=n_2$ and $U$ and $L$-nodes are
linked as in Figure~\ref{fig:corresponderMoreToBe}.

\begin{figure*}
\begin{center}
\mypic{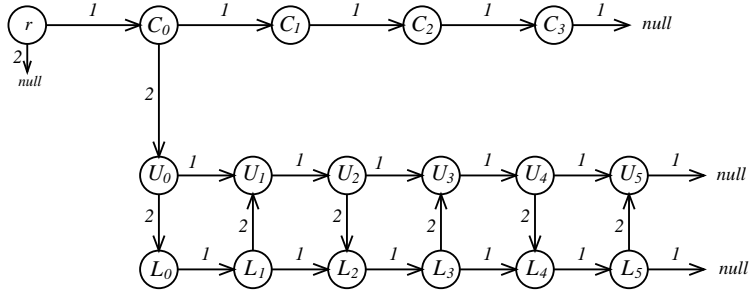}
\end{center}
\caption{Graph $G_0$ after establishing $s_2$ edges between
$U$ and $L$-nodes
\label{fig:corresponderMoreToBe}}
\end{figure*}

We next consider $s_2$-edges of $C$-nodes and find the
values $u_1,\ldots,u_{k-1}$ and $l_1,\ldots,l_{k-1}$.  First
we show that $s_2(C_{2j})$ is a $U$-node for $0 \leq j <
k$.  Suppose $s_2(C_{2j})$ is not a $U$-node.  Because $G_0
\homs P$, we conclude $s_2(C_{2j}) \notin V_0$.  But then
$G_0 \homs Q_{14}$, a contradiction.  We similarly establish
from $G_0 \homs P$ and $\lnot G_0 \homs Q_{14}$ that
$s_2(C_{2j+1})$ is an $L$-node for $0 \leq j < k$.  From
$G_0 \homs P$ it follows that $s_2(C_{2j})=U_{2i}$ for some
$i$ and $s_2(C_{2j+1})=L_{2f+1}$ for some $f$.  

We can therefore define $u_j$ and $l_j$ such that
\[\begin{array}{rcl}
  s_2(C_{2j})   &=& U_{2u_j} \\
  s_2(C_{2j+1}) &=& U_{2l_j+1} \\
\end{array}\]
for $0 \leq j < k$.

We next show
\[
   s_2(L_{2j})=\rootConst \ \ifandonlyif \
  \exists i\ s_2(C_{2i})=U_{2j}
\]
From $G_0 \homs P$ we have $s_2(L_{2j}) \in
\{\nullConst,\rootConst\}$.  Moreover, if
$s_2(C_{2i})=U_{2j}$, then $s_2(L_{2j})=\nullConst$.  It
remains to show that $s_2(L_{2j})=\rootConst$ implies
$s_2(C_{2i})=U_{2j}$ for some $i$.  Suppose that
$s_2(L_{2j})=\rootConst$ but $\forall i.\> s_2(C_{2i})\neq
U_{2j}$.  Then $G_0 \homs Q_{10}$, a contradiction.  We
similarly establish
\[
   s_2(U_{2j+1})=\rootConst \ \ifandonlyif \
  \exists i\ s_2(C_{2i+1})=L_{2j+1}
\]
using $\lnot\ G_0 \homs Q_{11}$.

We claim
\[
   u_j < u_{j+1}
\]
for $0 \leq j < k-1$.  Suppose the claim is false and let
$j$ be the smallest index for which $u_{j+1} \leq u_j$.  Let
$i$ be such that $u_i$ is the largest among
$u_0,\ldots,u_{k-1}$ with the property $u_i < u_{j+1}$.
Clearly $0 \leq i \leq j-1$.  Then there exists a homomorphism
$h$ from $G_0$ to $Q_{12}$ such that $h(U_{2u_i})=a_0$ and
$h(U_{2u_{j+1}})=a_4$.  This is a contradiction with $\lnot\
G_0 \homs Q_{12}$.  We similarly conclude
\[
   l_j < l_{j+1}
\]
for $0 \leq j < k-1$, using $\lnot\ Q_{13}$.

Finally we observe that we have identified all $s_2$-edges
from $V_0$, so $s_2[V_0] \subseteq V_0$.  Therefore $V
\setminus V_0 = \emptyset$.  We conclude that $G_0$ is
isomorphic to
\[
    \CG(n_1,k,u_1,\ldots,u_{k-1},l_1,\ldots,l_{k-1})
\]
\end{proof}

\subsection{The Undecidability Result}

\begin{theorem} \label{thm:implicationUndecidable}
The implication of graphs is undecidable over the class of
heaps.
\end{theorem}
\begin{proof}
We will reduce satisfiability of graphs over the class
of corresponder graphs to the problem of finding a
counterexample to an implication of graphs over the class of
heaps.  Given the reduction in
Proposition~\ref{prop:PCPreduction}, this will establish
that the implication of graphs is Turing co-recognizable and
undecidable.

Let $G$ be a graph.  Consider the implication
\begin{equation} \label{eqn:theImplication}
  (G \times P) \ \graphimplies_{\allheaps} \ Q
\end{equation}
We claim that $G_0$ is a counterexample for this implication
iff $G_0$ is a corresponder graph such that $G_0 \homs G$.

Assume that $G_0$ is a corresponder graph and $G_0 \homs G$.
By Proposition~\ref{prop:corrEncoding}, we have $G_0 \homs
P$ and $\lnot G_0 \homs Q$.  We then have $G_0 \homs (G
\times P)$.  Since $\lnot G_0 \homs Q$, we conclude that
$G_0$ is a counterexample for~(\ref{eqn:theImplication}).

Assume now that $G_0$ is a counterexample
for~(\ref{eqn:theImplication}).  Then $G_0 \homs G \times P$
and $\lnot G_0 \homs Q$.  Since $G_0 \homs P$ and $\lnot G_0
\homs Q$, by Proposition~\ref{prop:corrEncoding} we conclude that
$G_0$ is a corresponder graph.  Furthermore, $G_0 \homs G$.
\end{proof}

\subsection{Discussion}

In this section we give comments on our proof of the
undecidability of implication and state some implications of
this result for checking properties of programs.

\subsubsection{Graph Equivalence and Negation}

\begin{definition}
We say that graphs $G_1$ and $G_2$ are equivalent
over the class of graphs $C$ and write
\[
   G_1 \graphequiv_C G_2
\]
iff
\[
    G_0 \homs G_1 \ \ifandonlyif \ G_0 \homs G_2
\]
for every graph $G_0 \in C$.
\end{definition}

\begin{proposition}
Equivalence of graphs over the class of heaps is
undecidable.
\end{proposition}
\begin{proof}
From Proposition~\ref{prop:conjunction} we have
\[
    G_1 \graphimplies_{\allheaps} G_2
\]
iff
\[
    G_1 \graphequiv_{\allheaps} G_1 \times G_2
\]
The result then follows from
Proposition~\ref{thm:implicationUndecidable}.
\end{proof}

We also observe that regular graph constraints over heaps
are not closed under the negation.  Indeed, assume that for
every graph $G$ there exists a graph $\overline{G}$ such
that the heap models of $\overline{G}$ are all heaps that
are not models of $G$.  Then finding a counterexample to an
implication $P \graphimplies_{\allheaps} Q$ is reduced to
satisfiability of the graph
\[
    P \times \overline{G}
\]
This is a contradiction because
Proposition~\ref{prop:heapSatOk} implies that
satisfiability over heaps is decidable whereas
Proposition~\ref{thm:implicationUndecidable} implies that
finding a counterexample to $P \graphimplies_{\allheaps} Q$ is
undecidable.

\subsubsection{Implication of Acyclic Heaps} 
\label{sec:acyclicImplication}

Corresponder graphs are a cyclic subclass of the class of
heaps.  The cyclicity, however, is not at all essential for
our construction.  We argue that implication of graphs is
also undecidable over the class of acyclic heaps.  We can
define a minor variation of corresponder graphs where $U$
and $L$ nodes never point back to $\rootConst$.  Instead, we
introduce a special node different from $\nullConst$ to
indicate the difference between columns $j$ for
\[
   j \in \{u_1,\ldots,u_{k-1}\} \cup \{l_1,\ldots,l_{k-1}\}
\]
and the remaining columns.  The resulting graphs are acyclic
heaps.  
As a result, we have the following fact.
\begin{proposition}
Implication of graphs is undecidable over the class of acyclic
heaps.
\end{proposition}

\subsubsection{Alternative Proofs}

An alternative way to prove undecidability would be to show
that conjunction of regular graph constraints and their
negations can express graphs similar to grids (instead of
corresponder graphs).  While the construction using grids
may be possible, we have found the construction using
corresponder graphs to be simpler.  The reason is that
corresponder graphs, unlike grids, are essentially
one-dimensional structures.

Our proof of Proposition~\ref{prop:corrEncoding} could
potentially be simplified by showing that a larger fragment
of MSOL can be written in the form of negation of an
implication of graphs.  We consider formulas that can be
reduced to checking negation of an implication between
graphs.  Let a {\em literal} be a formula constructed from
an orable graph as in Section~\ref{sec:graphsEMSOL}.  Define
a {\em homogeneous clause} as a disjunction of positive
literals:
\[
   C_i = A^0_i \lor \cdots \lor A^{n_i}_i
\]
or a disjunction of negative literals:
\[
   D_i = (\lnot B^0_i) \lor \cdots \lor (\lnot B^{m_i}_i)
\]
Then any conjunction of positive and negative clauses
\[
    C_0 \land \cdots \land C_{n-1} \land 
    D_0 \land \cdots \land D_{m-1}
\]
is expressible as a negation of implication of graph
constraints.  This fragment appears quite expressive, but we
have not been able to obtain a characterization of the
fragment that allows a natural encoding a subclass like
grids or corresponder graphs in a way simpler than in
Proposition~\ref{prop:corrEncoding}.

\subsubsection{Consequences for Program Checking}
\label{sec:stepwiseInvs}

Implication of graphs arises if procedure specifications are
regular graph constraints.

\begin{example}

\newcommand{\procedure}{\m{procedure }}
\newcommand{\pre}{\m{pre }}

\begin{figure}
$\begin{array}{l}
   \procedure p() \\
   \pre G_1 \\
   \{ \\
   \qquad q(); \\
   \} \\
  \end{array}
$

$\begin{array}{l}
   \procedure q() \\
   \pre G_2 \\
   \{ \\
   \qquad \ldots \\
   \} \\
  \end{array}
$
\caption{Program Checking Requires $G_1 \graphimplies G_2$
\label{fig:procCallImplication}}
\end{figure}

Consider a procedure $p$ whose precondition is that the
program heap is homomorphic to a graph $G_1$ and a procedure
$q$ whose precondition is that the program heap is
homomorphic to a graph $G_2$
(Figure~\ref{fig:procCallImplication}).  If the first
statement in the body of $p$ is a call to $q()$, a program
checker must ensure that implication 
$G_1 \graphimplies G_2$ holds.
\end{example}

\noindent
We next show that the implication problem also arises when
maintaining an invariant at every program point, if the
invariant is a regular graph constraint.
\begin{figure}[thb]
\begin{center}
\mypic{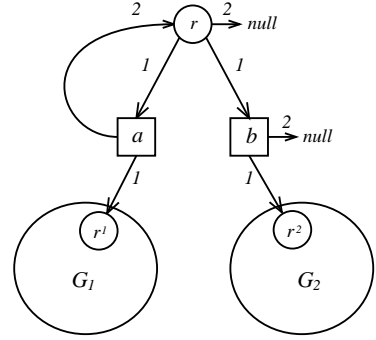}
\end{center}
\caption{Ensuring an invariant requires implication
\label{fig:stepwiseImplication}}
\end{figure}
Let
\[
   G^1 = \tu{V^1,s^1_1,s^1_2,\nullConst,\rootConst^1}
\]
\[
   G^2 = \tu{V^2,s^2_1,s^2_2,\nullConst,\rootConst^2}
\]
be orable graphs such that there are no edges from nodes
$V^1$ to $\rootConst^1$ and no edges from nodes $V^2$ to
$\rootConst^2$.  

\noindent
Construct the graph $G$ (Figure~\ref{fig:stepwiseImplication})
as
\[
   G = \tu{V,s_1,s_2,\nullConst,\rootConst}
\]
where
\[\begin{array}{rcl}
    V &=& \{\rootConst,a,b \} \cup V^1 \cup V^2 \\
    s_1 &=& \{ \tu{\rootConst,a}, \tu{\rootConst,b},
               \tu{a,\rootConst^1}, \tu{b,\rootConst^2} \} \\
        &\cup& s^1_1 \cup s^2_1 \\
    s_2 &=& \{ \tu{\rootConst,\nullConst},
               \tu{a,\rootConst}, \tu{b,\nullConst} \} \\
        &\cup& s^1_2 \cup s^2_2 \\
\end{array}\]
and where
\[
   \{\rootConst,a,b\} \cap (V^1 \cup V^2) = \emptyset
\]

\noindent
Suppose that we have a program checking system that verifies
that a graph constraint is true after every statement.
Consider the statement
\begin{equation} \label{eqn:assignmentStatement}
    \rootConst.1.2 := \nullConst
\end{equation}
Let
\[
   G^0 = \tu{V^0,s^0_1,s^0_2,\nullConst^0,\rootConst^0}
\]
be the graph before the statement.  After the statement
the resulting graph is
\[
   G^1 = \tu{V^0,s^0_1,s^1_2,\nullConst^0,\rootConst^0}
\]
where the value of $2$-edge from $x$ has changed
so that it points to $\nullConst$:
\[
   s^1_2 = s^0_2[x:=\rootConst]
\]
Our program checking system needs to verify that for all
heaps $G_0$,
\begin{equation} \label{eqn:originalImplication}
    (G^0 \homs G) \ \implies \ (G^1 \homs G)
\end{equation}

Let $h$ a homomorphism from $G^0$ to
$G$.  Let $x = s^0_1(s^0_1(\rootConst^0))$.
Then
\[
   h(x) = \rootConst^1
\]
or
\[
   h(x) = \rootConst^2
\]
Moreover, $\rootConst^1$ and $\rootConst^2$ are reachable
only through the path $\rootConst^0,1,a,1$, so no nodes other
than $x$ may be mapped to $\rootConst^1$ or $\rootConst^2$.
We can therefore show that the
implication~(\ref{eqn:originalImplication}) is equivalent
to
\begin{equation}
   G_1 \graphimplies G_2
\end{equation}
As explained in Section~\ref{sec:acyclicImplication}, we can
modify the construction in the proof of
Proposition~\ref{prop:corrEncoding} such that $P$ and $Q$
have no edges terminating at $\rootConst$.  We then let
$G^1 = P$ and $G^2=Q$.  From the undecidability of the
implication of graphs over the domain of heaps it follows
that maintaining an invariant expressed as a regular graph
constraint is undecidable, even across a simple assignment
statement such as~(\ref{eqn:assignmentStatement}).

%% file: related.tex
\section{Related Work}

The idea of typestate as system for statically verifying
changing properties of objects was proposed
in~\cite{StromYemini86Typestate} and extended
in~\cite{StromYellin93ExtendingTypestate}.  The original
typestate system as well as the more recent work in the
context object oriented
programming~\cite{DrossopoulouETAL01Reclassification} do not
support constraints over dynamically allocated objects,
which is the focus of our paper.

Several recent systems support tree-like dynamically
allocated data
structures~\cite{SmithETAL00AliasTypes,
WalkerMorrisett00AliasTypesRecursive,
FradetMetayer97ShapeTypes,Moeller01PALE}.  The restriction
to tree-like data structures is in contrast to our notion of
heap, which allows cycles.  The presence of non-tree data
structures is one of the key factors that make the
implication of regular graph constraints undecidable.

The idea of representing properties of a statically
unbounded number of heaps by homomorphically mapping them to
a bounded family of graphs is pervasive in the work on shape
analysis~\cite{LarusHilfinger88DetectingConflictsAccesses,
ChaseETAL90AnalysisPointersStructures,
GhiyaHendren96TreeOrDag, SagivETAL96Destructive,
SagivETAL99Parametric}.  These analyses use abstractions
that capture approximate properties of data structures even
if they are not tree-like.  This feature of shape analyses
makes our results directly applicable.  Our undecidability
result implies inability to semantically check implication
or equivalence of such abstractions.  

Shape analysis techniques were applied to a typestate
checking problem in role
analysis~\cite{KuncakETAL02RoleAnalysis}.  The
compositionality of the analysis and the presence of
procedure specifications made the need for solving the
implication of constraints
in~\cite{KuncakETAL02RoleAnalysis} explicit.  The
algorithm~\cite{KuncakETAL02RoleAnalysis} uses ``context
matching'' as a decidable approximation for the implication
of constraints.  In~\cite{Kuncak01DesigningRoleAnalysis} it
was suggested that the implication problem for role
constraints is undecidable.  The argument makes use of
acyclicity constraints as well as the constraints on the
number of incoming edges of a node.  In the present paper,
we have shown that undecidability holds even for the regular
graph constraints, which cannot directly specify acyclicity
or the number of incoming edges of a node.  This makes the
present undecidability result strictly stronger than the
result in~\cite{Kuncak01DesigningRoleAnalysis}.

We were pleased to discover that the constraints derived as
a simplification of role analysis constraints generalize the
notions of tree
automata~\cite{Thomas97LanguagesAutomataLogic,
ComonETAL97Tata} and a whole family of equivalent systems
over
grids~\cite{GiammarresiRestivo97TwoDimensionalLanguages}.
The remarkable fact that MSOL over trees is equivalent to
tree automata inspired the question which classes of graphs
have decidable MSOL theory~\cite{Courcelle97MSOL}.  In this
paper we have introduced regular graph constraints which can
be seen as a alternative to MSOL in generalizing projections
of local properties over trees and grids.  Although regular
graph constraints are strictly weaker than MSOL (and in fact
the satisfiability of regular graph constraints is decidable
over heaps), we have shown that the implication for regular
graph constraints over heaps is undecidable.

\nocite{BoergerETAL97ClassicalDecisionProblem}

%% file: conclusion.tex
\section{Conclusion}

We have proposed regular graph constraints as an abstraction
of mutually recursive properties of objects in potentially
cyclic graphs.  We presented some evidence that regular graph
constraints are a natural generalization of the tree automata
and domino systems.  We have shown that satisfiability of
regular graph constraints is decidable over the domain of
heaps.  As a main result, we have shown that the implication
of regular graph constraints is undecidable.  The
consequence of this result is that verifying that procedure
preconditions are satisfied as well as maintaining program
invariants is undecidable if these properties are expressed
as regular graph constraints.

We have seen that decidability of problems with regular
constraints is sensitive to the choice of the class of
graphs.  In particular, a smaller class of graphs need not
imply better decidability properties.  This indicates that
techniques for reasoning about different classes of graphs
may be substantially different.  We conclude that a good
support for mechanized reasoning about data structures would
likely contain a set of specialized reasoning techniques for
different classes of graphs.

%% file: acknowledge.tex
\paragraph{Acknowledgements}
We thank 
Chandrasekhar Boyapati,
Yuri Gurevich, 
Patrick Lam
and
Andreas Podelski
for useful discussions.
We thank Chandrasekhar Boyapati and Patrick Lam for useful
comments on a draft of this paper.